\documentclass[a4paper,10pt]{article}
\usepackage[utf8]{inputenc}
\usepackage[english]{babel}
\usepackage{ifthen}
\usepackage{calc}

\usepackage{times}      
\usepackage{mathptmx}   
\usepackage{amsmath,amsfonts,amssymb}
\usepackage{graphicx,xcolor}
\usepackage{booktabs}
\usepackage{authblk}
\usepackage{multirow}

\usepackage[normalem]{ulem}

\usepackage[
backend=biber,
style=nature,
]{biblatex}
\usepackage[left=2.5cm,%
right=2.5cm,%
top=2.5cm,%
bottom=2.5cm,%
headheight=12pt,%
letterpaper]{geometry}%

\usepackage[labelfont={bf,sf},%
labelsep=period,%
justification=raggedright]{caption}

\usepackage[colorlinks=true, allcolors=blue]{hyperref}
\usepackage{placeins}
\usepackage{float}
\usepackage{setspace}

\usepackage{todonotes}

\title{Atomic Cluster Expansion Potentials for Screw Dislocations in BCC Refractory Metals}
\author[]{Lei Zhang\thanks{l.zhang9@tue.nl} }

\author[]{Francesco Maresca\thanks{f.maresca@rug.nl}}
\affil[]{Mechanics of Materials, Engineering and Technology Institute Groningen, Faculty of Science and Engineering, University of Groningen, Nijenborgh 4, 9747 AG Groningen, The Netherlands}

\date{}

\begin{document}
\setstretch{1.5}
\flushbottom
\maketitle
\thispagestyle{empty}

\begin{abstract}
Accurate atomistic modeling of screw dislocations in body-centered cubic (bcc) metals remains challenging because their plasticity is governed by a complex dislocation glide behavior due to their compact three-fold symmetric core structure and a strongly temperature-dependent flow stress induced by the large Peierls barrier. In the context of group 6 (V, Nb, Ta) and group 5 (Mo, W) refractory metals (RMs), both classical interatomic potentials and some machine learning potentials consistently fail to reproduce density functional theory (DFT) Peierls barriers and the glide plane. Here, we developed an array of atomic cluster expansion (ACE) potentials for these RMs by extending an existing DFT database. The developed ACE potentials significantly improve the description of screw dislocation properties, achieving near-DFT accuracy for Mo and W and substantial improvement for V, Nb, and Ta. The results show that transferability to screw dislocation behavior depends sensitively on both database composition and element-specific energetics, and that achieving a single-humped Peierls barrier alone is not a sufficient validation metric for accurate prediction of dislocation glide. For Nb, Mo, and W, the developed ACE models also enable reliable calculation of kink-pair activation enthalpies, which are well described by both Kocks' law and a line-tension model.
\end{abstract}

\section{\label{sec:intro}Introduction}

Body-centered cubic (bcc) refractory metals (RMs) and their alloys combine high strength at elevated temperatures with generally poor tensile ductility and limited fracture toughness at low temperatures~\cite{ashby1985influence}. A characteristic feature of these systems is the brittle-to-ductile transition (BDT) that occurs with increasing temperature~\cite{johnson1962ductile,han2022mechanism}. The limited plastic deformation at low temperatures originates from the low mobility of screw dislocations ($1/2\langle 111\rangle$), which in bcc RMs is governed by thermally activated mechanisms~\cite{argon2007strengthening,caillard2003thermally}. In contrast, non-screw dislocation characters exhibit much higher mobility~\cite{christian1983some,duesbery1991dislocation}. The low mobility of screw dislocations in RMs is attributed to their unique core structure; the non-planar core introduces significant lattice resistance~\cite{vitek1970core,ito2001atomistic}. Consequently, extensive research has focused on understanding the connection between screw dislocation structure and plastic flow in RMs~\cite{vitek1970core,wen2000atomistic,maresca2018screw,dezerald2016plastic,starikov2025dislocation}.

Among the available approaches, atomistic modeling has proven to be an indispensable tool for investigating the structure and energetics of screw dislocations. Density functional theory (DFT) is widely used to calculate the core structure, Peierls barrier, and core trajectory of $1/2\langle 111\rangle$ screw dislocations~\cite{ventelon2013ab,dezerald2014ab,dezerald2016plastic,weinberger2013peierls}. However, the high computational cost prevents its application at finite temperatures and strain rates, which motivates the use of classical molecular dynamics (MD). Although being less accurate, MD offers a substantial computational speedup and can treat systems containing millions of atoms, enabling the study of dislocation interactions and finite temperature behavior. The accuracy of MD depends critically on the quality of the interatomic potentials (IAPs). Classical IAPs, such as the embedded atom method (EAM) and angular dependent potential (ADP), often predict artificial core structures and inaccurate Peierls potentials~\cite{gordon2010screw,weinberger2013peierls,cereceda2013assessment,starikov2024angular}. These limitations have driven the development of machine learning interatomic potentials (ML-IAPs) tailored to dislocations~\cite{goryaeva2021efficient,wang2022classical,zhang2024efficiency,yin2021atomistic,li2020complex}.

To date, several ML-IAPs have been proposed for bcc RMs~\cite{byggmastar2020gaussian,nikoulis2021machine,goryaeva2021efficient,wang2022classical,kwon2023accurate,ito2024machine,pan2024atomic,shuang2025modeling} and for refractory high entropy alloys (RHEA)~\cite{yin2021atomistic,lyu2023effects,li2020complex,jiang2026scaling,luo2026dislocation,wang2025ductility,byggmastar2026nine}. While some of the ML-IAPs were developed to model dislocation behavior, they still lack comprehensive validation against dislocation properties, especially those of screw dislocations. For example, moment tensor potential (MTP) and spectral neighbor analysis potential (SNAP) were developed to investigate dislocation mobility in MoNbTaW RHEA~\cite{yin2021atomistic,li2020complex}. However, neither SNAP nor MTP is capable of predicting the DFT Peierls barrier of screw dislocations in Nb, Mo, and W~\cite{starikov2024angular}. Another representative example is a family of ``general purpose'' Gaussian approximation potentials (GAP) constructed for V, Nb, Ta, Mo, and W (denoted here as GAP-2020)~\cite{byggmastar2020gaussian}. GAP-2020 has been shown to accurately reproduce a broad range of properties, including elasticity, point defects, surface energies, thermal expansion, phonons, and melting point. However, we show in the following that GAP-2020 fails to predict the Peierls potential for RMs. This limitation highlights that good agreement with bulk and surface properties does not necessarily guaranty transferability to dislocations that are essential to simulate plastic deformation. 

\begin{figure}[h]
	\centering
	\includegraphics[width=0.8\linewidth]{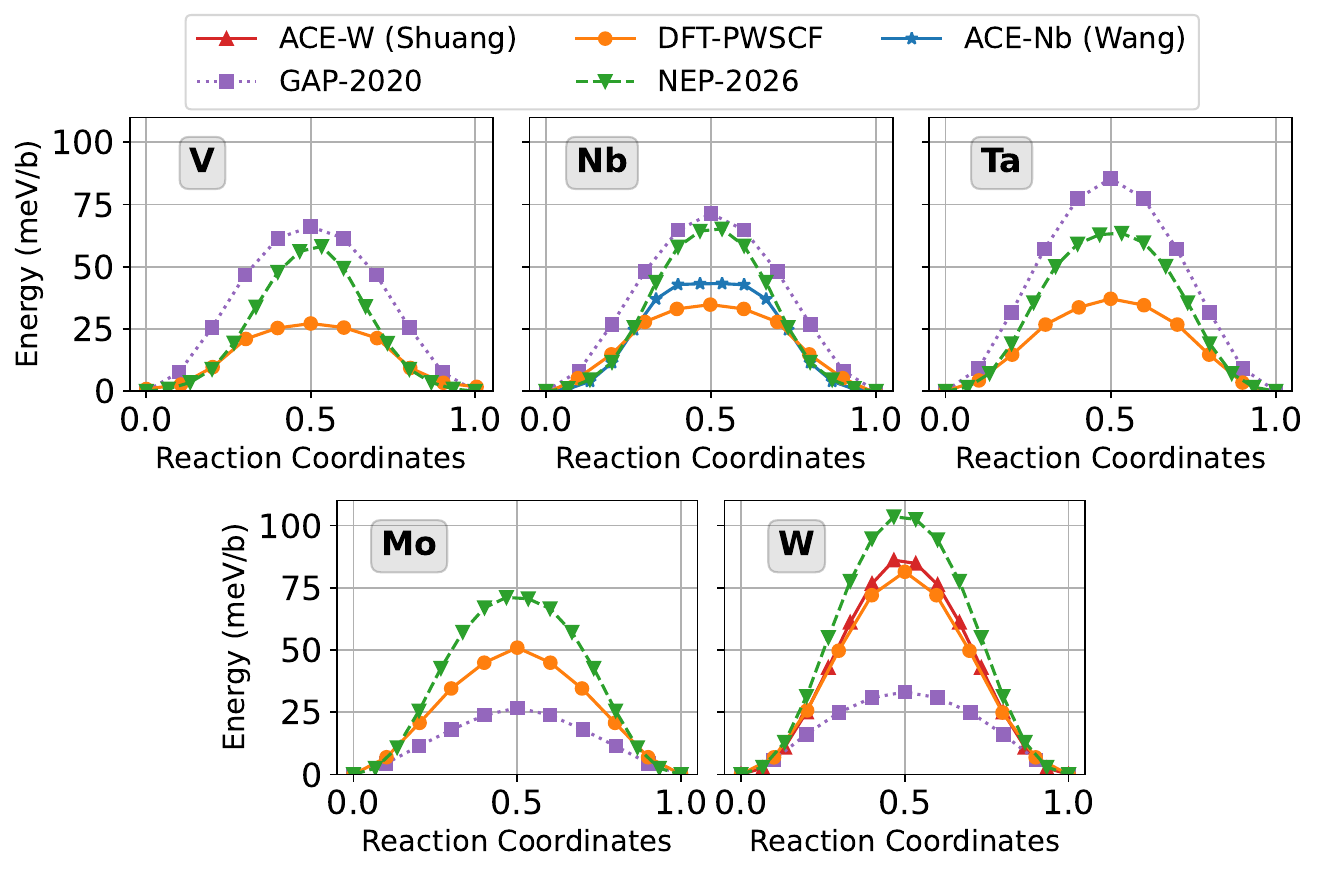}
	\caption{Peierls barriers of screw dislocations in V, Nb, Mo, Ta, and W. GAP-2020 predictions were obtained using the original GAP models~\cite{byggmastar2020gaussian}. DFT-\textit{PWSCF} data were taken from Ref.~\cite{dezerald2014ab}. ACE-2025-Shuang data were taken from Ref.~\cite{shuang2025universal}. ACE-2025-Wang data were taken from Ref.~\cite{wang2025ductility}. NEP-2026 data were taken from Ref.~\cite{byggmastar2026nine}.}
	\label{fig:gap_predictions}
\end{figure}

To further assess the state-of-the-art of ML potentials, we compute the Peierls potentials of five RMs, including V, Nb, Ta, Mo, and W, using the GAP-2020~\cite{byggmastar2020gaussian} and NEP-2026~\cite{byggmastar2026nine} models, as shown in Fig.~\ref{fig:gap_predictions}. In addition, we calculate the Peierls potential using recently reported potentials for Nb and Ta from the NbTaTiHf ACE potential~\cite{wang2025ductility} and for W from Ref.~\cite{shuang2025modeling}. For Ta, however, the nudged elastic band (NEB) calculation could not be converged using the ACE potential~\cite{wang2025ductility}. DFT data taken from Ref.~\cite{dezerald2014ab} are referred to as DFT-PWSCF. GAP-2020 and NEP-2026 predict a single-hump Peierls barrier for all RMs. The Peierls barriers predicted by GAP-2020 deviate from the DFT results by more than a factor of two. For the group 5 elements, GAP-2020 tends to overestimate the Peierls barriers, whereas for the group 6 elements, it underestimates them. NEP-2026 overestimates the Peierls barriers for all RMs. In summary, none of the existing ML-IAPs, in their current form, are accurate enough to describe screw dislocation behavior for five elementary RMs.

A previous study on W showed that directly including screw dislocation configurations in the DFT database improved the accuracy of both the Nye tensor and Peierls barrier predictions~\cite{szlachta2014accuracy}. Moreover, it has been shown that systematic optimization of ML-IAPs, with respect to both the ML framework and the model hyperparameters, can significantly improve both accuracy and efficiency~\cite{lysogorskiy2021performant,zhang2024efficiency}. In this work, we extend the existing DFT database~\cite{byggmastar2020gaussian} and train ML-IAPs within the Atomic Cluster Expansion (ACE) framework. The ACE framework provides a systematically improvable polynomial basis, a linear parametrization that enables robust regularization, and high computational efficiency suitable for large scale dislocation simulations~\cite{drautz2019atomic,lysogorskiy2021performant}. ACE naturally supports simple and inexpensive measures of model extrapolation, which can be used as practical proxies for model uncertainty~\cite{lysogorskiy2023active}. This is particularly valuable when the configurations relevant to the target application are not explicitly represented in the training database.

The remainder of the paper is organized as follows. Section~\ref{sec:method} describes the computational methods, model parameters, and training and testing procedures. Section~\ref{sec:validation} presents the validation of bulk and surface properties. The screw dislocation properties predicted by the developed models are discussed in Section~\ref{sec:Dislocations}. Section~\ref{sec:kp_enthalpy} examines the kink-pair activation enthalpy barrier in Nb, Mo, and W, and compares the atomistic results with theoretical models. Section~\ref{sec:discussion} discusses the Peierls barriers in comparison with multiple DFT references and existing IAP predictions, and examines the potential impact of the lattice rescaling strategy used to construct the DFT databases. Finally, Section~\ref{sec:conclusion} summarizes the main conclusions of the paper. 

\section{\label{sec:method}Methods}
\subsection{\label{sec:method-dft}DFT calculations}
In this work, the GAP-2020 database developed in \cite{byggmastar2020gaussian} is considered as a starting point to build an enriched DFT database. Consistent with \cite{byggmastar2020gaussian}, all DFT calculations for the enriched database are performed using the Vienna Ab initio Simulation Package (VASP)~\cite{kresse1993ab,kresse1994ab,kresse1996efficient,kresse1996efficiency}. Perdew-Burke-Ernzerhof (PBE) generalized-gradient approximation exchange correlation functional \cite{perdew1996generalized} was employed. Hard projector-augmented wave (PAW) pseudopotentials were used, which are kept the same as the original database (V\_sv, Nb\_sv, Ta\_pv, Mo\_sv,  W\_sv)~\cite{byggmastar2020gaussian}. The energy cutoff of the plane-wave expansion is set to 500 eV. First order Methfessel-Paxton method with 0.1 eV smearing is applied. $k$-points were sampled on Monkhorst-Pack grids with a spacing of 0.15 \AA$^{-1}$.  Energy and forces of the new configurations are included when training the new potentials. The number of newly added configurations and their structural categories are listed in Appendix~\ref{append:dft_data}. The complete DFT database is available in the \textit{Data Availability} section.

The original GAP-2020 database includes distorted unit cells, high-temperature structures, point defects, flat and disordered surfaces, liquids, and a range of crystal structures (sc, bcc, fcc, hcp, A15, C15)~\cite{byggmastar2020gaussian}. The database was initially developed for W and then rescaled according to the lattice parameters of V, Nb, Ta, and Mo. GAP-2020 is developed for general purposes, focusing on radiation damage simulations, with no validation against dislocation properties. Our calculations show that GAP-2020 is unable to capture quantitatively the Peierls barrier of screw dislocations in all five RMs considered here (Fig.~\ref{fig:gap_predictions}). To improve the description of screw dislocations, we first enrich the GAP-2020 DFT database with highly distorted primitive-cell configurations, following an approach previously used for crack-related applications in bcc Fe~\cite{zhang2023atomistic}. These configurations expand the range of local atomic environments of the GAP-2020 database (see Appendix~\ref{append:pca_analysis} for the principal component analysis). In total, 2,963 primitive cells were generated for each element by rescaling iron primitive cells obtained via SOBOL sampling to the element-specific lattice parameters.

The enrichment based on the primitive cell sampling is already sufficient to enable the prediction of the screw dislocation Peierls barrier in group 6 elements (Mo,W). For group 5 elements (V, Nb, Ta), further enrichment of the database is required to reproduce the Peierls potential of the screw dislocations. For these elements, we additionally include the equations of state for the bcc, fcc, and hcp structures~\cite{wang2024taming,aitken2024controlling}, as well as stacking fault configurations on the \{110\} and \{112\} planes. Finally, in preparation for future fracture relevant applications, we also include traction-separation (T-S) data for all elements. The T-S curves were generated by separating the bulk material along the \{100\} and \{110\} surfaces for all elements and include both equilibrium and rattled configurations. The rattled configurations were obtained by adding random Gaussian perturbations to the atomic Cartesian coordinates.

In this work, nudged elastic band (NEB) calculations using VASP are performed to obtain the Peierls barrier of screw dislocation for five RMs~\cite{henkelman2000climbing}. A screw dislocation dipole configuration is used for the NEB calculations, and its exact geometry is described in Section~\ref{sec:method-mc_md}. Both dislocations are moved to their next Peierls valley in the NEB final configuration. The initial and final configurations are fully relaxed before the NEB calculation. Nine intermediate replicas are generated along the NEB path by linearly interpolating between the initial and final configurations. During the NEB calculations, the force tolerance is set to $10^{-2}$ eV/Å and the NEB inter-replica spring constant is set to the default VASP value $k=-5$ eV/Å$^2$.

\subsection{\label{sec:method-database} ACE model Training}
The ACE formalism introduces a complete description of multi-body atomic interactions \cite{drautz2019atomic}. The energy of atom $i$ can be expressed as a function of multiple per-atom atomic properties $\psi_i^{p}$
\begin{equation}
    E_i = \mathcal{F}(\psi^{(1)}_i,\psi^{(2)}_i,...,\psi^{(P)}_i)\ .
\end{equation}
We employ complex embeddings that can lead to a more flexible description of the energy landscape at a slightly higher computational cost~\cite{erhard2024modelling}
\begin{equation}
    E_i = (\psi^{(1)}_i)^{1/4} + (\psi^{(2)}_i)^{1/2} + (\psi^{(3)}_i)^{3/4} + \psi^{(4)}_i + (\psi^{(5)}_i)^{5/4} + (\psi^{(6)}_i)^{3/2} + 
    (\psi^{(7)}_i)^{7/4} + (\psi^{(8)}_i)^{2}\ ,
\end{equation}
where atomic property $\psi^{(p)}_i$ is calculated based on the local atomic environments:
\begin{equation}
    \psi^{(p)}_i = \sum_{\mathbf{\nu}} c_{\mathbf{\nu}}^{(p)} \mathbf{B}_{i\mathbf{\nu}} \ ,
\end{equation}
where $c_{\mathbf{\nu}}^{(p)}$ is the expansion coefficients and $\mathbf{B}_{i\mathbf{\nu}}$ is the basis functions that depend on the atomic positions and species. 

\begin{figure}[h!!!]
	\centering
	\includegraphics[trim=0 0 0 0, clip, width=0.8\linewidth]{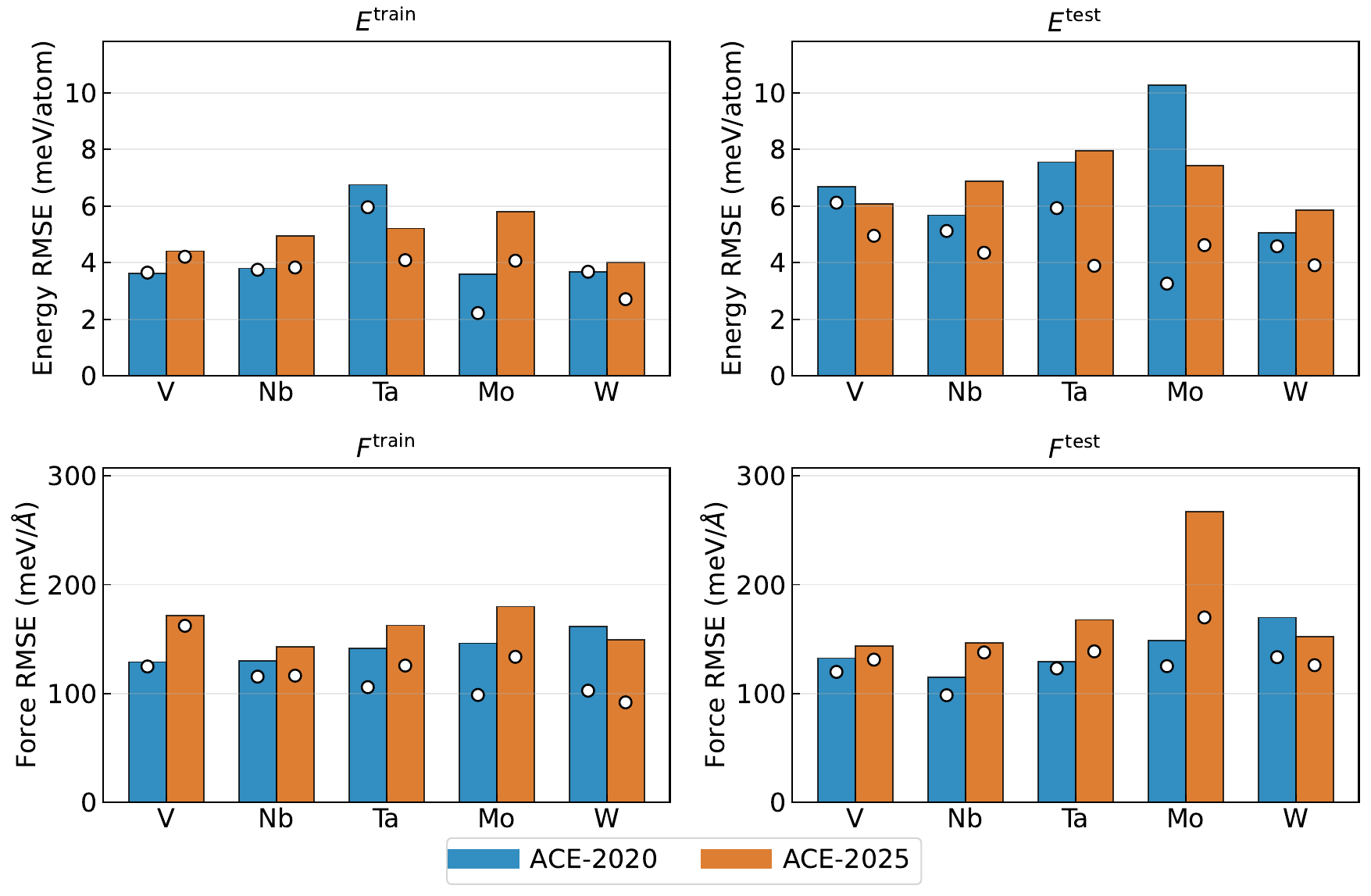}
	\caption{
    Comparison of the energy and force RMSEs of the ACE-2020 and ACE-2025 potentials for V, Nb, Ta, Mo, and W. The upper panels show the training and test energy RMSEs, while the lower panels show the corresponding force RMSEs. Solid bars denote the RMSEs evaluated on the full dataset, and open circles indicate the RMSEs of datasets inside 1 eV/atom energy convex hull. 
		} 
	\label{fig:ace_rmse}
\end{figure}

ACE potentials are trained using \textit{PACEMAKER}~\cite{lysogorskiy2021performant,bochkarev2022efficient}. 
Two versions of the ACE potential, denoted ACE-2020 and ACE-2025, are constructed using the original and enriched DFT databases, respectively. 
This comparison allows us to assess how the quality and diversity of the DFT training data influence the accuracy of the resulting ACE potentials. 
As shown in Fig.~\ref{fig:ace_rmse}, all training and testing energy RMSE values for ACE-2025 are below 8~meV/atom, demonstrating the improved accuracy obtained with the enriched database. 
Training on the original database is particularly challenging for Ta; therefore, we use the optimization package \textit{Xpot} to identify a parameter set that yields a reasonably low RMSE~\cite{thomas2024hyperparameter}.

In addition to the RMSE-based validation, we also evaluate the reliability of the ACE potentials during atomistic simulations. 
Our previous work highlighted the importance of uncertainty quantification in plasticity simulations~\cite{zhang2024efficiency}. 
\textit{PACEMAKER} implements a D-optimality criterion based on the MaxVol algorithm, which quantifies the extrapolation grade $\gamma$ of new atomic environments~\cite{podryabinkin2017active,lysogorskiy2023active}. 
Here, the extrapolation grade $\gamma$ is monitored during the simulations to assess whether the sampled atomic environments remain within the interpolation regime of the training database, thereby providing an additional measure of simulation reliability beyond the training and testing RMSEs.

\subsection{\label{sec:method-mc_md}Molecular Statics (MS) calculations}

\begin{figure}[h!!!]
	\centering
	\includegraphics[trim=0 0 0 0, clip, width=0.6\linewidth]{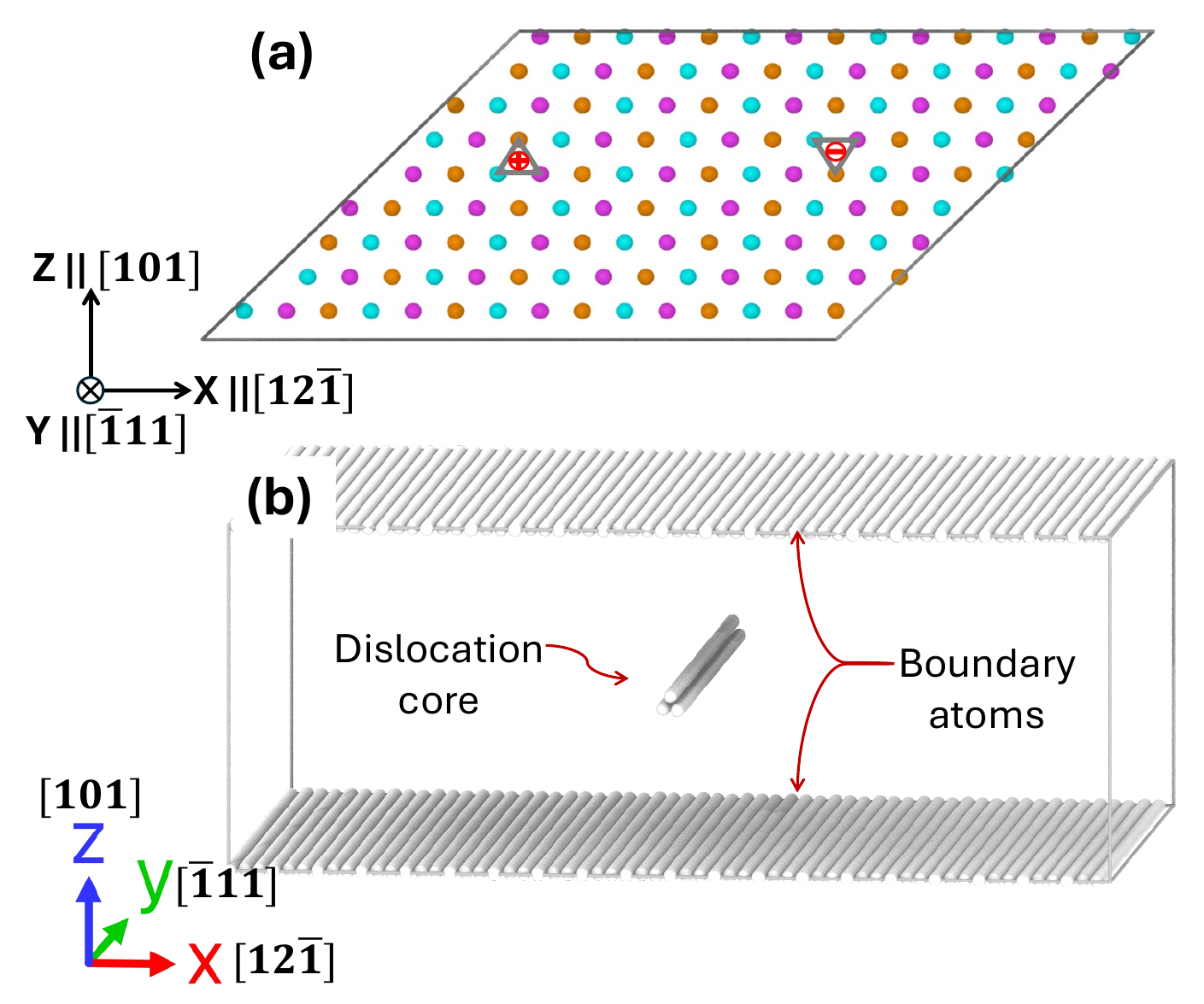}
    \caption{
    Schematic representations of the screw-dislocation configurations. 
    (a) Screw-dislocation dipole configuration. The dislocation core positions are indicated by $\oplus$ and $\ominus$ symbols inside the triangles, and circles represent atomic positions. Colors denote the atomic positions in the pristine bcc crystal for one periodic unit cell of length $b$ along the dislocation-line direction $\gamma$: orange atoms are used as the reference position ($0$), pink atoms are located at $b/3$, and cyan atoms at $2b/3$. The crystallographic directions are also indicated. 
    (b) Periodic array of dislocations (PAD) configuration, showing the free-surface atoms and the dislocation core. The structure is analyzed using the adaptive common neighbor analysis (CNA) method available in OVITO~\cite{stukowski2009visualization}. Atoms identified as bcc are removed for clarity. The crystallographic directions along $x$, $y$ and $z$ are shown. 
}
	\label{fig:schematic_plot}
\end{figure}

A dislocation dipole configuration consisting of two parallel screw dislocations with opposite Burgers vectors is used to calculate the Peierls potential, as shown in Fig.~\ref{fig:schematic_plot}a~\cite{li2004core}. The simulation cell contains 135 atoms. The crystal orientations along the $x$, $y$, and $z$ directions are set to $[12\bar{1}]$, $[\bar{1}11]$, and $[101]$, respectively. The resulting Burgers vector, $1/2\langle 111 \rangle$, and slip plane, $(1\bar{1}0)$, correspond to the dominant slip system in bcc metals. Volterra dislocations, assumed to be infinitely long and perfectly straight, are introduced into an anisotropic elastic medium by imposing the displacement field obtained from anisotropic elasticity theory~\cite{stroh1958dislocations}. The dislocation dipole configuration is stable and allows a significant reduction in the simulation cell size, which makes this approach particularly useful for \textit{ab initio} modeling of dislocations.
Periodic boundary conditions are applied in all directions. A quadrupolar periodic array of dislocation dipoles is generated by tilting the simulation box by half a periodicity vector along the slip direction $x\|[12\bar{1}]$~\cite{ventelon2013ab}. With this modification, the dislocations have alternating Burgers vectors along both in plane directions ($x$--$y$ plane), which leads to zero net stress at the dislocation centers through the nearly symmetric superposition of the elastic fields~\cite{cai2003periodic}. The nudged elastic band (NEB) calculations of the Peierls potential are then carried out using this dipole configuration.

We make use of the Python package \textit{atomman} to generate the dislocation dipole configurations~\cite{atomman}. The Atomic Simulation Environment (ASE)~\cite{larsen2017atomic} and \textit{matscipy}~\cite{grigorev2024matscipy} were used to perform NEB calculations with the ACE potentials. Nine intermediate replicas were generated by linear interpolation between the initial and final configurations. The force tolerance was set to $10^{-3}$ eV/\AA, and the NEB spring constant between adjacent replicas was set to $k=0.1$ eV/\AA$^2$. The scripts used to generate screw dislocations, differential displacement (DD) maps, and perform the NEB calculations are described in the Data Availability section. Atomistic configurations were visualized using OVITO~\cite{stukowski2009visualization}.

For the calculation of the kink-pair nucleation barrier under an applied shear stress, a periodic array of dislocations (PAD) configuration was employed, as shown in Fig.~\ref{fig:schematic_plot}b. The simulation cell has dimensions $(l_x = 60a) \times (l_y = 40b) \times (l_z = 40c)$ and contains 96,000 atoms. Here, $a=\sqrt{2/3}a_0$ is the spacing between adjacent Peierls valleys, $b=\sqrt{3}/2\,a_0$ is the magnitude of the Burgers vector, $c=\sqrt{2}/2\,a_0$ is the interplanar spacing of the $\{101\}$ planes, and $a_0$ is the lattice parameter. Periodic boundary conditions were applied along the slip direction $[12\bar{1}]$ and the dislocation line direction $[\bar{1}11]$. All atomistic models were generated using LAMMPS (version \textit{10 Dec 2025})~\cite{thompson2022lammps}, and the NEB calculations were also performed in LAMMPS. We use 32 replicas and the FIRE algorithm, with a force tolerance of $5\times 10^{-2}$ eV/\AA, for the NEB calculations~\cite{bitzek2006structural}.

\section{\label{sec:results}Results}
\subsection{\label{sec:validation}Bulk and surface property validations}
\begin{figure}[h!!!]
	\centering
	\includegraphics[width=0.7\linewidth]{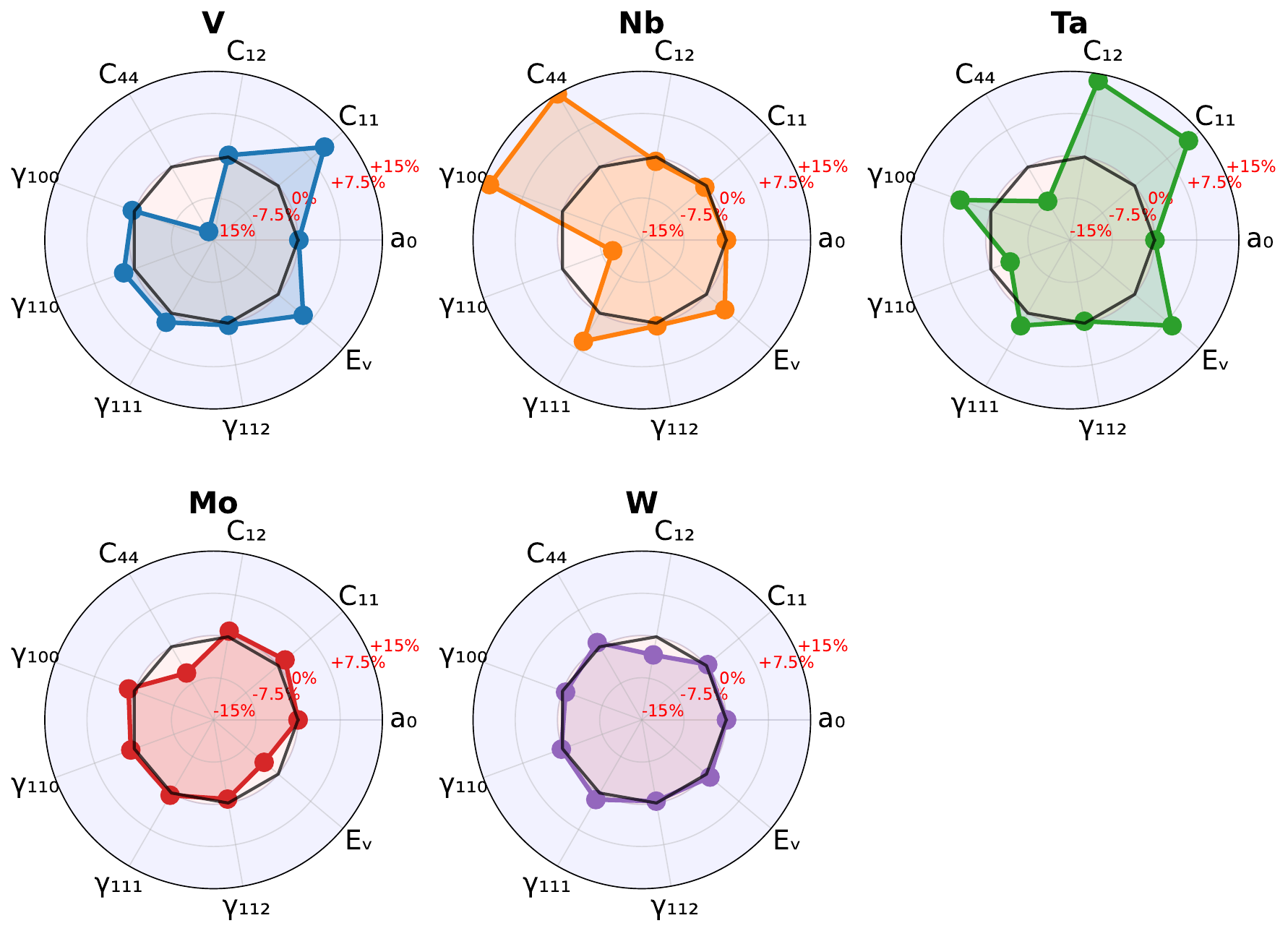}
	\caption{Assessment of the ACE-2025 model accuracy across fundamental properties of bcc refractory metals. The relative error (\%) for each property, including lattice parameter ($a_0$), elastic constants ($C_{ij}$), surface energies for the $\{100\}$, $\{110\}$, $\{111\}$, and $\{112\}$ crystallographic planes, and the mono-vacancy formation energy, is shown for V, Nb, Ta, Mo, and W. The error is calculated with respect to DFT values from Ref. \cite{byggmastar2020gaussian}.}
	\label{fig:basic_properties}
\end{figure}

GAP-2020 potentials were originally developed for elastic, thermal, liquid, defect, and surface properties. It has been shown that they accurately predict these properties (e.g. equation of state, phonon dispersion, melting curve) \cite{byggmastar2020gaussian}. Based on ACE-2025, we predict the lattice parameter ($a_0$), the vacancy formation energy ($E_v$), the elastic constants ($C_{11}$, $C_{12}$, and $C_{44}$), the surface energies of low-index planes ($\gamma_{100}$, $\gamma_{110}$, $\gamma_{111}$, and $\gamma_{112}$). Fig.~\ref{fig:basic_properties} summarizes the performance of ACE-2025 by showing the relative errors with respect to DFT reference values, which are taken from Ref.~\cite{byggmastar2020gaussian}. The relative error is defined as $\left|\frac{p^{\rm ACE}-p^{\rm DFT}}{p^{\rm DFT}}\right| \times 100\%$, where $p$ is the property. For Mo and W, the relative errors of all considered properties remain below 5\%, indicating an excellent level of agreement with DFT. The overall performance is further quantified using the average error index $Q$ proposed in Ref.~\cite{zhang2024efficiency}, obtaining $Q=$ 3.66\%, 5.55\%, 6.04\%, 1.58\%, and 0.86\% for V, Nb, Ta, Mo, and W, respectively. These values show that the ACE-2025 models are particularly accurate for Mo and W, while the errors for V, Nb, and Ta are somewhat larger but still within a reasonable range for high-fidelity large-scale simulations.

\begin{figure}[h!!!]
	\centering
	\includegraphics[width=0.7\linewidth]{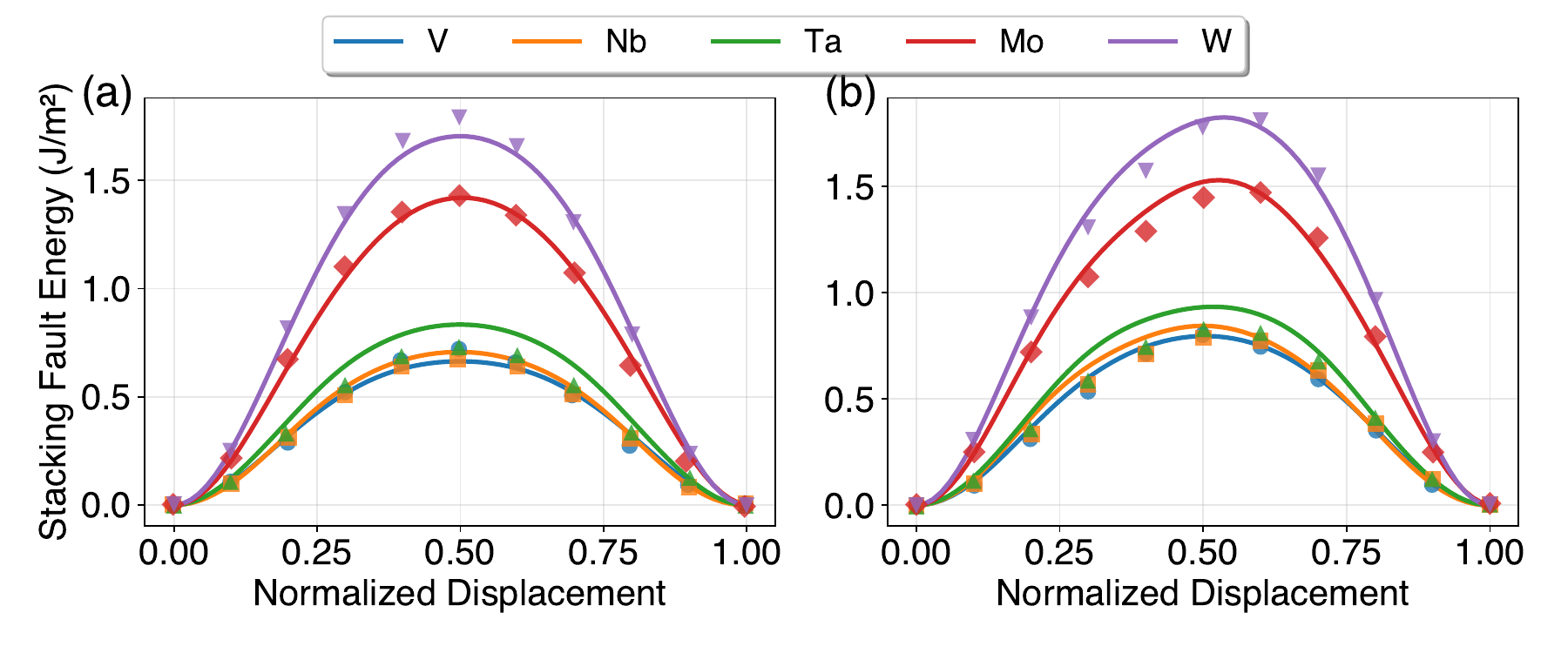}
	\caption{Generalized stacking fault energy (GSFE) curves for five bcc RMs predicted by ACE-2025 and DFT \cite{zhang2023unstable}. (a) $\langle 111\rangle$\{110\} and (b)$\langle 111\rangle$\{112\} slip system.}
	\label{fig:stacking_fault}
\end{figure}

We evaluate the generalized stacking fault (GSF) energies on the two primary slip planes, \{110\} and \{112\}, using the ACE-2025 interatomic potentials for V, Nb, Ta, Mo, and W. Fig. \ref{fig:stacking_fault} compares the resulting GSF energy profiles (solid lines) with DFT reference data (symbols) taken from Ref.~\cite{zhang2023unstable}. For all five RMs and on both slip planes, the ACE-2025 potentials reproduce the characteristic shape of the GSF curves. In particular, ACE-2025 is able to capture the asymmetry behaviour of \{112\} plane. The ACE-2025 prediction for Ta slightly deviates from the DFT results. 

\subsection{\label{sec:Dislocations}Screw dislocation core structure and trajectory}

\begin{figure}[h!!!]
	\centering
	\includegraphics[width=0.8\linewidth]{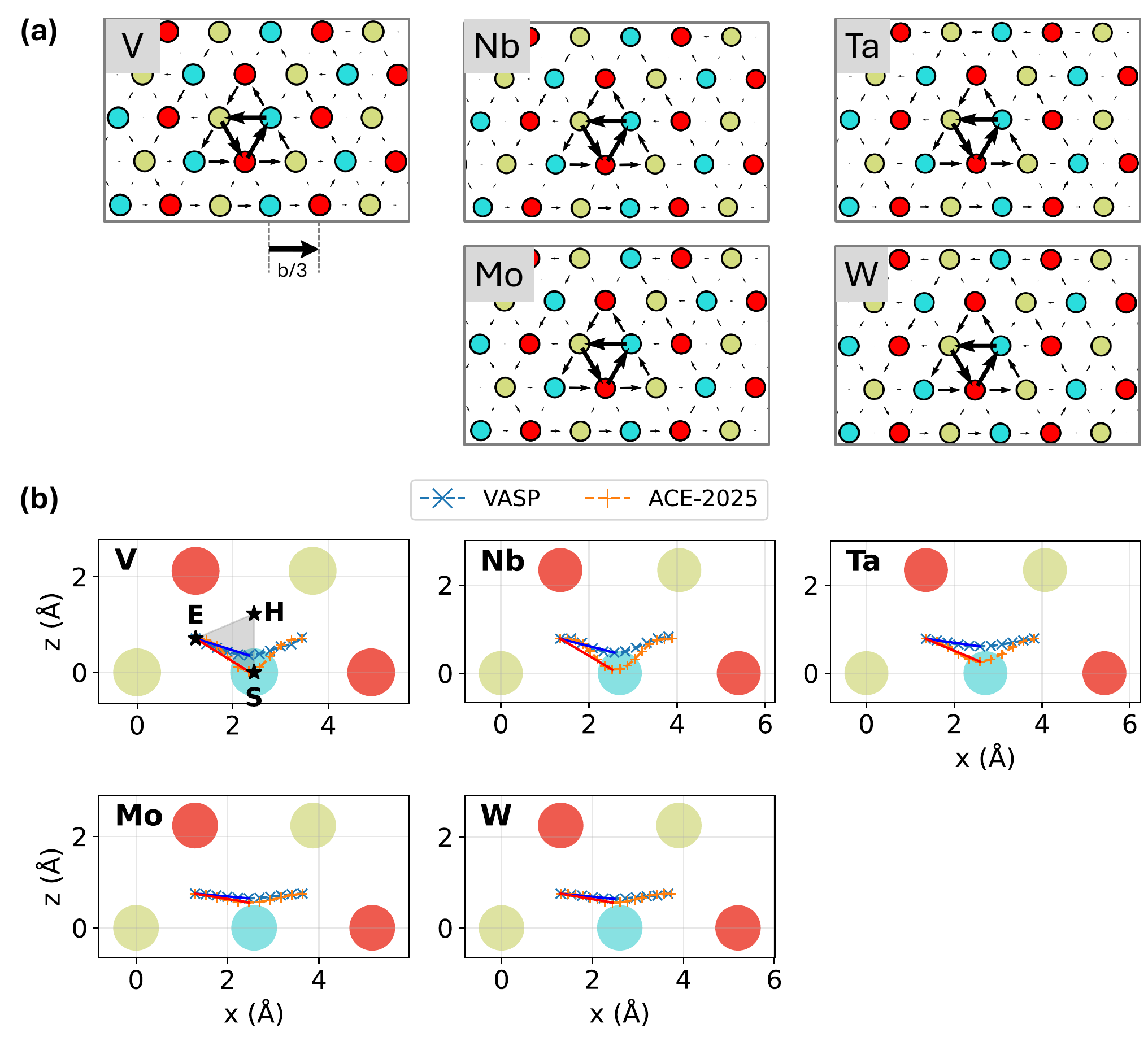}
	\caption{(a) $\langle 111 \rangle$-zone differential displacement maps  at the dislocation core for five elements. Atomic positions are represented by the circles. Atoms are colored by their height along out-of-the-plane (dislocation line direction $y$) $\langle 111 \rangle$ in the undeformed perfect state for one periodic unit cell of length $b$. Red atoms are used as the reference position ($0$), green atoms are located at $b/3$, and cyan atoms at $2b/3$. 
    Arrows are calculated from the relative displacements before and after introducing the dislocation. The arrows are normalized by $b/3$, which is equal to the spacing of neighboring atoms. 
    (b) Dislocation-core migration trajectories along the Peierls barrier between easy-core configurations in bcc metals. 
    The deviation angle $\alpha$ is defined between the blue/red solid line and the horizontal direction, where the blue/red solid line connects the lowest-energy point along the trajectory to the easy-core position. 
    The triangular region in the V panel indicates the minimum-energy paths connecting the easy-core (E), hard-core (H), and split-core (S) configurations.
    }
	\label{fig:ddm}
\end{figure}

Differential displacement maps (DDMs) were obtained by comparing the displacement along $z$ directions before and after introducing the screw dislocation. Fig. \ref{fig:ddm}a is the $\langle 111\rangle$-zone DDM of the relaxed dislocation core structure at $T=0$ K and $\tau=0$ MPa. Atoms are colored according to their height along $z$ directions in the undeformed state. Red, blue, and yellow represent three different layers along $\langle 111 \rangle$ direction. DDMs reveal that ACE-2025 potentials predict non-degenerate core geometry for five elements, in agreement with DFT predictions \cite{dezerald2014ab}. 

\begin{figure}[h!!!]
	\centering
	\includegraphics[width=0.9\linewidth]{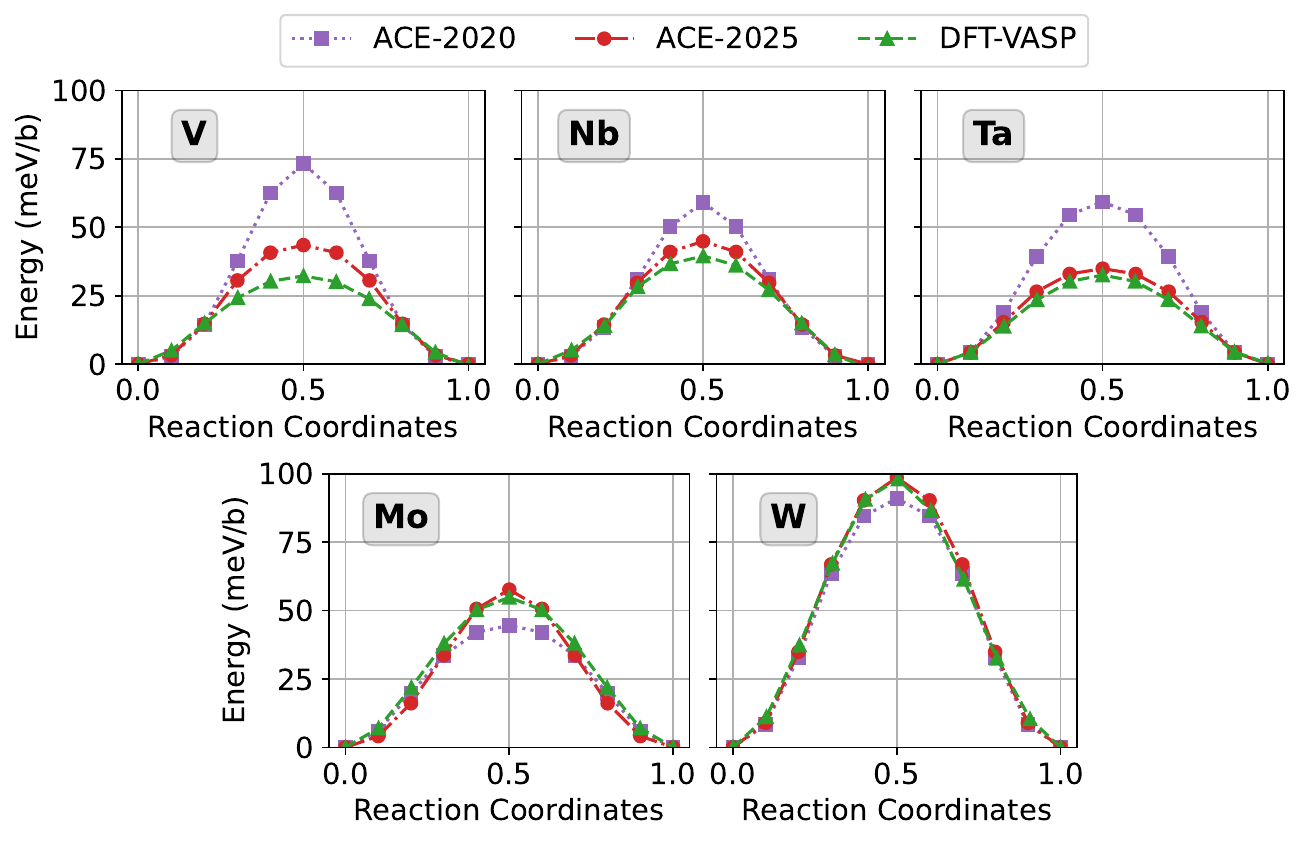}
	\caption{Screw dislocation Peierls potential at $T = 0$~K and zero applied stress in bcc V, Nb, Ta, Mo, and W. Results obtained with the ACE-2020 and ACE-2025 interatomic potentials are compared to our VASP calculations.}
	\label{fig:Peierls_barrier_all}
\end{figure}

\begin{table}[h!!!]
    \centering
		\renewcommand{\arraystretch}{1.2}
    \caption{Deviation angle $\alpha$ (\textdegree) of the screw dislocation trajectory along Peierls potential and the relative position of the saddle point $\xi^{\rm saddle}_{\rm H2S}$ along the hard-to-core (H2S) migration pathway in bcc V, Nb, Ta, Mo, and W. Results from ACE-2025 are compared with DFT calculations from Refs.~\cite{dezerald2014ab,dezerald2016plastic} and the present study.}
    \begin{tabular}{c|c|ccccc}
    \hline
         & Method & V & Nb & Ta & Mo & W  \\
    \hline
	\multirow{3}{*}{$\alpha$ (\textdegree)}	
      & DFT~\cite{dezerald2016plastic} & -10.6 & -21.7 &  -7.1 & -9.9 & -10.9\\
     & DFT & -17.58  & -14.42 &  -8.01 & -4.88 & -5.5\\   
     & ACE-2025 & -31.82 & -29.17 & -20.45 & -9.24 & -9.13 \\
     \hline     
	\multirow{2}{*}{$\xi^{\rm saddle}_{\rm H2S}$}	
     & DFT~\cite{dezerald2014ab}  & 0.63 & 0.68 & 0.51 & 0.57 & 0.55 \\
     & ACE-2025 & 0.71 & 0.67 & 0.73 & 0.67 & 0.62  \\
     \hline
    \end{tabular}
    \label{tab:core_traj}
\end{table}
We calculate the Peierls potentials for V, Nb, Ta, Mo and W using VASP (Fig. \ref{fig:Peierls_barrier_all}). ACE-2020 can predict the Peierls barrier in W with error less than 10 meV/b, while errors in Mo and Nb are less than 20 meV/b. However, the Peierls barriers for V and Ta are significantly overestimated compared to the DFT-VASP reference. Incorporating primitive cell configurations into the training database did not resolve this issue, and the revised potential still could not quantitatively capture the Peierls barriers of V, Nb, and Ta. To improve the description of screw dislocation core energetics, we enriched the database with more stacking fault configurations and equations of state for multiple crystal structures (bcc, fcc, and hcp)~\cite{wang2024taming}. In particular, extra screw dislocation configurations from NEB calculations are added to the database of Nb. Based on the expanded database, the ACE-2025 potentials reproduce the Peierls barriers of Nb and Ta with high accuracy, with errors of only 5.40 and 2.30~meV/$b$, respectively, while only slightly overestimating the barrier in V by 11.30~meV/$b$. The remaining discrepancy for vanadium may stem from the rescaling strategy in constructing the initial database, which will be discussed in Section \ref{sec:discussion}. The extrapolation-grade distribution $\gamma<1$ of ACE-2025 along the NEB paths indicate that the sampled configurations lie within the interpolation regime of the potential (Fig.~\ref{fig:gamma_neb}).

The dislocation core trajectory along the Peierls potential pathway is highly relevant for the plastic anisotropy behavior in bcc metals, including the twinning/antitwinning (T/AT) asymmetry, deviations from Schmid's law \cite{dezerald2016plastic}, and the tendency to slip on glide planes other than the $\{110\}$ with the maximum resolved shear stress. The dislocation core trajectory can be evaluated via either the cost function approach or the stress method~\cite{ventelon2013ab,dezerald2014ab}. Here, the dislocation core trajectory is evaluated using the stress approach~\cite{dezerald2014ab,dezerald2016plastic}. Based on the NEB results, the stress variation is measured for the replicas along the reaction pathway. The position of the dislocation is assumed to vary linearly between the initial and final configurations. The $z$ component of the distance between two dislocation core positions (Fig.~\ref{fig:schematic_plot}) is 
\begin{equation}
    \Delta z = \frac{S}{b} \left[
    \frac{(C_{11}-C_{12})\sigma_{xy} - C_{15}(\sigma_{xx}-\sigma_{zz})}
    {(C_{11}-C_{12})C_{44} - 2C_{15}^2}
    \right]\ ,
\end{equation} 
where $S$ is the area of the simulation cell perpendicular to the dislocation lines, and $C_{ij}$ is the elastic constants in Voigt notation in the crystallographic orientation of dislocation~\cite{chaari2014first}.
Fig.~\ref{fig:ddm}(b) compares the screw dislocation core trajectories obtained from VASP and from the ACE-2025 potentials. The corresponding deviation angles are also reported in Table~\ref{tab:core_traj}. For Nb, Mo, and W, the ACE-2025 predictions are in good agreement with the DFT results of Ref.~\cite{dezerald2016plastic}. By contrast, for V and Ta, the deviation angles are systematically overestimated, which may be attributed to an incomplete description of the energy landscape connecting the different core configurations (easy, hard, and split).

\begin{figure}[h!!!]
	\centering
	\includegraphics[width=0.9\linewidth]{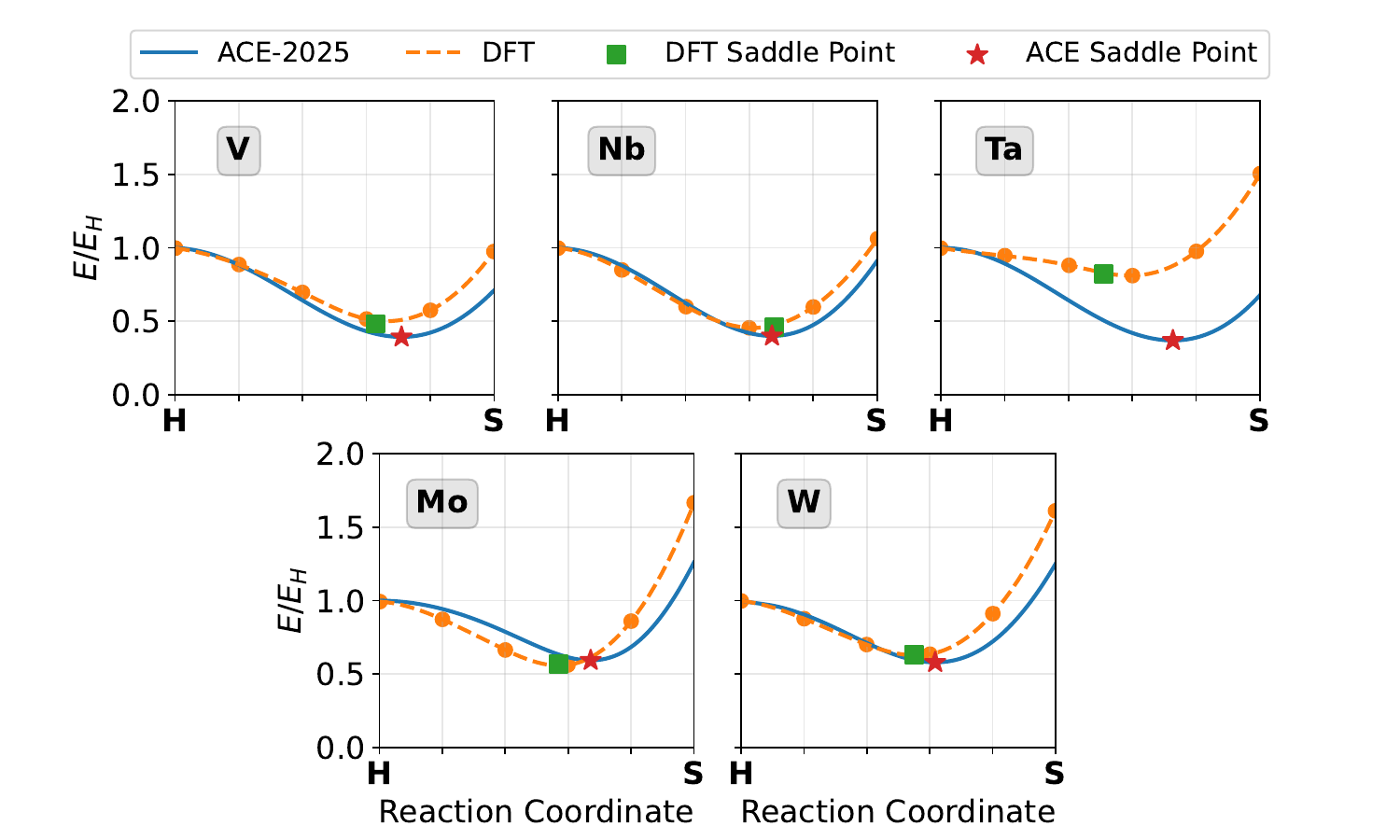}
	\caption{Energy variation between the hard and split cores of screw dislocation for five RMs. The energies are normalized by the minimum energy along the path. The dot and star symbols correspond to the saddle point along pathway predicted by DFT \cite{dezerald2014ab} and ACE-2025, respectively.}
	\label{fig:h2s_profile}
\end{figure}

To further analyze these discrepancies, we compute the energy pathway from the hard to the split core using the ACE-2025 potentials, as shown in Fig.~\ref{fig:h2s_profile}. DFT predicts $E_H \approx E_S$ for V and Nb~\cite{dezerald2014ab}; therefore, ACE-2025 underestimates the split-core energy in V. For Ta, ACE-2025 yields $E_H$ significantly higher than $E_S$, whereas DFT predicts the opposite ordering. In contrast, for W and Mo, the ACE-2025 models correctly reproduce the higher energy of the split core ($E_S$) relative to the hard core ($E_H$), consistent with DFT. The saddle points $\xi^{{saddle}}_{{H2S}}$ obtained from ACE-2025 and DFT are highlighted by stars and squares, respectively, in Fig.~\ref{fig:h2s_profile}. As summarized in Table~\ref{tab:core_traj}, ACE-2025 systematically overestimates $\xi^{{saddle}}_{{H2S}}$ for all elements, even in Mo and W, where the correct energy hierarchy ($E_S > E_H > E_P$) is captured, and in Nb, where $E_S\sim E_H$ according to DFT. For Ta, the large discrepancy in the saddle point might be traced back to the incorrect energy hierarchy among different core structures.

To summarize, ACE-2025 achieves near DFT accuracy in describing the screw dislocation energy landscape of Nb, Mo, and W. The Peierls potentials are well captured for all five elements. However, a fully quantitative reproduction of the more complex energy landscapes in V and Ta remains challenging. A plausible explanation is related to limitations and biases in the underlying DFT database used for training, which will be discussed in detail in Section~\ref{sec:discussion}. Moreover, although the Peierls potential is often used as a key benchmark for assessing the ability of IAPs to describe dislocation properties in bcc RMs~\cite{goryaeva2021efficient,starikov2021optimized,starikov2025dislocation}, an accurate Peierls barrier alone does not guarantee a faithful description of the relative energies of different screw dislocation core geometries, nor correct glide behaviour as it will be discussed next. 

\subsection{\label{sec:disloc_glide}Dislocation glide behavior in Nb, Mo, and W}

The analysis of the screw-dislocation two-dimensional Peierls potential and the corresponding core trajectories indicates that the ACE-2025 potentials for Nb, Mo, and W can faithfully reproduce the DFT-predicted core energetics and migration paths. However, these validations are performed for an infinitely long, perfectly straight dislocation line, whereas screw dislocations glide at low temperatures through kink-pair nucleation and propagation. To determine the enthalpy barrier for dislocation glide, the glide plane must usually be specified \textit{a priori} in energy barrier calculations. To avoid artificially constraining the dislocation motion to a \{110\} slip plane, we therefore perform MD simulations of screw-dislocation glide to identify the activated slip plane in Nb, Mo, and W.

To investigate the finite-temperature glide mechanism, we use a simulation cell of dimension $(l_x=60a)\times(l_y=52b)\times(l_z=30c)$, containing $N=93,600$ atoms. A shear stress is applied to the top and bottom boundary atoms in opposite directions (details of the force application are given in the next section). The system is first equilibrated for 20 ps in the NPT ensemble at $T=200~\mathrm{K}$. An initial shear stress of $\tau_{\rm init}=0.5\tau_{\rm P}$ is then applied, followed by an additional 200 ps of equilibration in the NVT ensemble. This step confirms that the dislocation remains immobile under $\tau_{\rm init}=0.5\tau_{\rm P}$. After equilibration, the applied shear stress is increased stepwise until dislocation motion is observed, using increments of 50 MPa every 200 ps. This corresponds to a loading rate of $2.5\times10^{8}~\mathrm{GPa\,s^{-1}}$.

\begin{figure}[h!!!]
	\centering
	\includegraphics[width=0.8\linewidth]{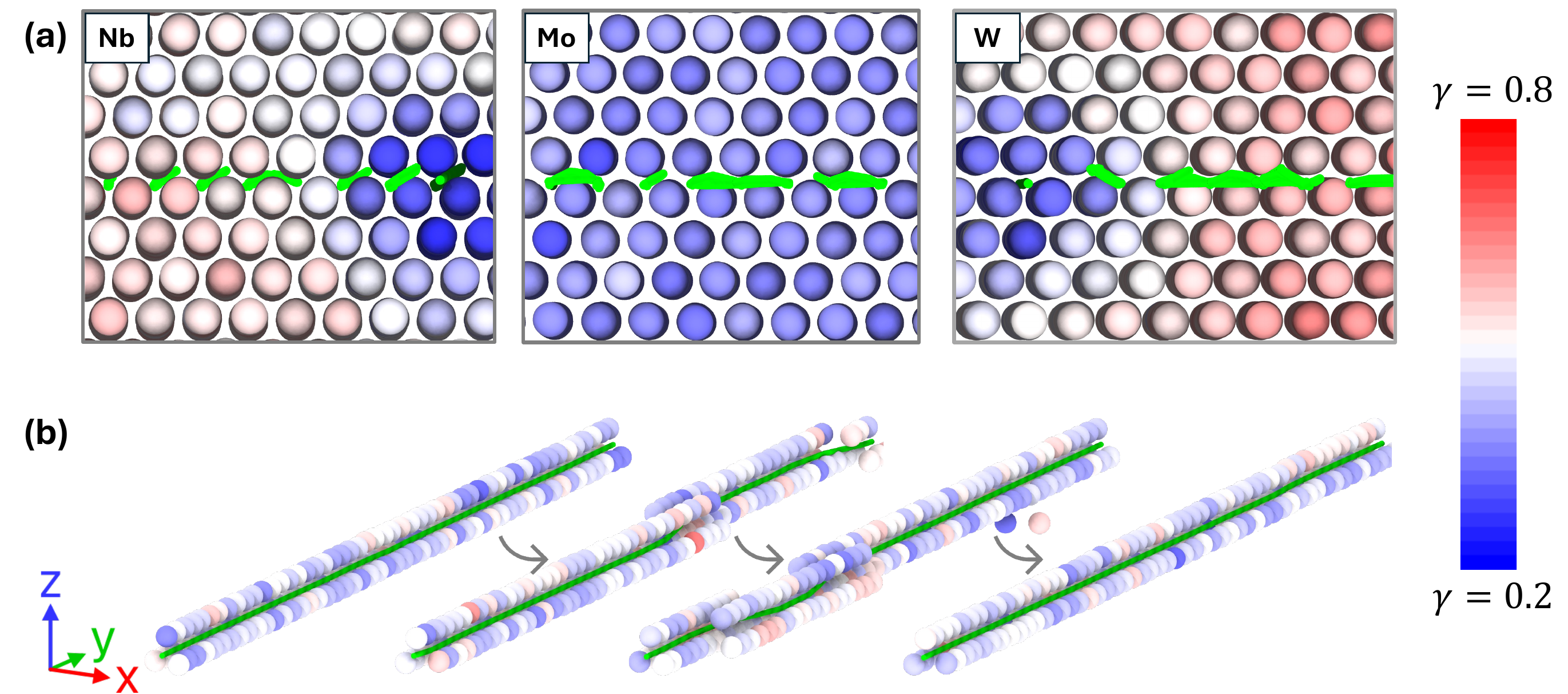}
	\caption{(a) Screw dislocation core trajectories in Nb, Mo, and W obtained from MD simulations at $\tau_{\rm{app}}=0.7, 1.2, 2.05$~GPa and $T=200~\rm{K}$. Green dots/lines represent sequential dislocation positions that change with simulation time. Dislocation positions are analysed using the DXA algorithm available in OVITO. 
    (b) Kink-pair nucleation and migration process in Nb.
    Atoms are colored according to the extrapolation grade $\gamma$.}
	\label{fig:dislocation_Nb_200K}
\end{figure}

Fig.~\ref{fig:dislocation_Nb_200K}(a) shows the dislocation-core trajectories of Nb, Mo, and W obtained from MD simulations at $T=200~\mathrm{K}$. The simulations reveal that the dislocations glide on the \{110\} planes. The trajectory fluctuations further indicate a strong tendency for cross slip in Nb, which is not observed in Mo or W. This behavior is consistent with the large deviation angle predicted for Nb (Table~\ref{tab:core_traj}). Previous calculations have also shown that screw dislocations can cross slip onto different \{110\} slip planes~\cite{zotov2021molecular,starikov2025dislocation}.

At finite temperature, screw-dislocation motion proceeds through kink-pair nucleation followed by lateral kink migration, as shown in Fig.~\ref{fig:dislocation_Nb_200K}. The atoms are colored according to the extrapolation grade $\gamma$ predicted by ACE-2025. Although fluctuations in $\gamma$ are observed during the simulation, the values remain below 1, indicating that the sampled configurations remain within the interpolation regime of the potential (see Fig.~\ref{fig:gamma_finiteT}).

Moreover, the twinning/anti-twinning (T/AT) asymmetry associated with glide on the $\{112\}$ plane is a characteristic feature of plastic anisotropy in bcc metals and is closely linked to non-Schmid effects~\cite{dezerald2016plastic}. Capturing this asymmetry is therefore essential for IAPs intended for large-scale simulations of plastic flow. Here, we evaluate screw-dislocation glide on the $\{112\}$ plane in both the T and AT directions at $T=0$~K.

Table~\ref{tab:crss_112_t_at} summarizes the critical resolved shear stress (CRSS) for glide in the two directions. For all three RMs, $\tau_{\mathrm{c}}^{\mathrm{AT}}$ is higher than $\tau_{\mathrm{c}}^{\mathrm{T}}$, confirming the expected T/AT asymmetry. To quantify this effect, we define the asymmetry factor $\chi=(\tau_{\mathrm{c}}^{\mathrm{AT}}-\tau_{\mathrm{c}}^{\mathrm{T}})/\tau_{\mathrm{c}}^{\mathrm{T}}$, and compare the ACE-2025 predictions with the DFT-based estimates reported in Ref.~\cite{dezerald2016plastic}. Mo and W exhibit only modest asymmetry ($\chi<0.4$), whereas Nb shows a much stronger asymmetry ($\chi>1.2$). ACE-2025 reproduces this element-dependent trend well, as shown in Table~\ref{tab:crss_112_t_at}.

In summary, ACE-2025 predicts screw-dislocation glide on the $\{110\}$ plane in Nb, Mo, and W at $T=200$~K, while maintaining bounded extrapolation grades throughout the simulations. It also reproduces the DFT trend in T/AT asymmetry on the $\{112\}$ plane. In the next section, we use ACE-2025 to evaluate the kink-pair nucleation enthalpy.

\begin{table}[h]
    \centering
    \renewcommand{\arraystretch}{1.3}
    \caption{Critical resolved shear stress (CRSS) on the $\{112\}$ slip plane in the twinning (T) and anti-twinning (AT) directions for Nb, Mo, and W.}
    \begin{tabular}{ccccc}
        \toprule
        \multirow{2}{*}{Element}
        & \multicolumn{2}{c}{$\tau_{\mathrm{c}}$ (GPa)}
        & \multicolumn{2}{c}{$\chi$} \\
        \cmidrule(lr){2-3} \cmidrule(lr){4-5}
        & T & AT & ACE-2025 & DFT estimate \\
        \midrule
        Nb & 1.25 & 3.75 & 2 & $ > 1.2$ \\
        Mo & 1.6 & 2.2 & 0.38 & 0.21 \\
        W  & 3.35 & 4.3 & 0.28 & 0.24 \\
        \bottomrule
    \end{tabular}
    \label{tab:crss_112_t_at}
\end{table}

\subsection{\label{sec:kp_enthalpy}Kink-pair nucleation enthalpy}
Screw dislocation motion is governed by a thermally activated process of kink-pair formation and migration. Here, the kink-pair nucleation enthalpy as a function of the applied shear stress is evaluated via NEB calculations and compared with theoretical models.
The rate associated with a thermally activated process is a key component in determining a dislocation mobility law. The kink-pair (KP) nucleation and propagation process can be described by the transition state theory (TST), i.e.
\begin{equation}
    R \propto \exp{\left(-\frac{\Delta H}{k_b T}\right)}\ ,
\end{equation}
where $R$ is the KP nucleation rate, $\Delta H$ is the activation enthalpy, $k_b$ is the Boltzmann constant, and $T$ is the absolute temperature. The activation enthalpy $\Delta H (\tau,T)$ is a function of the applied stress $\tau$ and temperature. Within the harmonic approximation, the temperature dependence is neglected~\cite{vineyard1957frequency}, and we therefore approximate the activation enthalpy only as a function of the applied stress, i.e. $\Delta H(\tau)$.

We first compute the enthalpy as a function of applied stress from the NEB calculation. To perform the NEB calculations, we used a PAD configuration (Fig.~\ref{fig:schematic_plot}b). A shear stress $\tau$ is applied to both the top and bottom boundary atoms in opposite directions during the NEB optimization. The stress is applied by adding forces to boundary atoms with a thickness of $6c$ (five layers of atoms). The force $\mathbf{f}$, corresponding to applied stress $\tau$, is computed as $\mathbf{f}(\tau) = \tau l_x l_y / N_{\rm atoms}$. Note that $\mathbf{f}(\tau)$ is the force exerted on each atom, where $N_{\rm atoms}$ is the total number of atoms in the top or bottom slab where forces are being applied. 

The enthalpy of replica $i$, with the reaction coordinate $\xi^i$, is computed as
\begin{equation}
	H(\tau,\xi^i) = E_p(\tau,\xi^i) - W^{\rm ext}(\tau,\xi^i),
\end{equation}
where $E_p(\tau,\xi^i)$ is the potential energy of replica $i$. In the atomistic calculations, $W^{\rm ext}(\tau,\xi^i)$ is the work done by the applied force $\mathbf{f}(\tau)$ on the boundary atoms,
\begin{equation}
	W^{\rm ext}(\tau,\xi^i) =
	\sum_{j \in \partial \Omega}
	\mathbf{f}_j(\tau) \cdot \mathbf{u}_j(\xi^i).
\end{equation}
Here, $j$ indicates the boundary atoms on which external forces are applied, and $\mathbf{u}_j(\xi^i)$ is the displacement vector of boundary atom $j$ in replica $i$. Note that both top and bottom boundary atoms need to be considered in this calculation, as shown in Fig.~\ref{fig:schematic_plot}. The work can already be subtracted in LAMMPS by combining `\textit{fix addforce}' and `\textit{fix\_modify energy}' commands. Example scripts are provided in the \textit{Data Availability} Section. 

\begin{table}[h!!!]
    \centering
		\renewcommand{\arraystretch}{1.2}
    \caption{Kink-pair formation enthalpy at zero applied stress, $\Delta H^\ast(0)$, and Peierls stress, $\tau_P$, for Nb, Mo, and W. ACE-2025 predictions are compared with values obtained from experimental fits, line-tension (LT) models parameterized using DFT Peierls potentials, and previous DFT calculations. The Peierls stresses reported for ACE-2025 were obtained from direct MS simulations at $T=0$~K.}
    \begin{tabular}{c|c|ccccc}
    \hline
         & Method & Nb & Mo & W  \\
    \hline
	\multirow{3}{*}{$\Delta H^\ast (0)$}
     & Expt. & 0.62~\cite{seeger2006slip}  & 1.19~\cite{hollang1997flow} & 2.05~\cite{brunner2000comparison} \\
      & LT based on DFT~\cite{dezerald2016plastic} & 1.28 & 1.05 & 1.54 \\
     & ACE-2025 & 0.72 & 0.99 & 1.49\\
     \hline     
	\multirow{4}{*}{$\tau_P$}	
     & Exp. & 0.45~\cite{seeger2006slip} & 0.87~\cite{hollang1997flow} & 0.9~\cite{brunner2000comparison} \\
     & DFT~\cite{dezerald2014ab} (Disregistry) & 0.86 & - & 1.8 \\
     & DFT~\cite{weinberger2013peierls} & 0.74 & 1.6 & 2.4 \\
     & ACE-2025 & 1.375 & 1.525  & 2.875\\
     \hline
    \end{tabular}
    \label{tab:kink_pair_energy}
\end{table}

\begin{figure}[h!!!]
	\centering
	\includegraphics[width=0.6\linewidth]{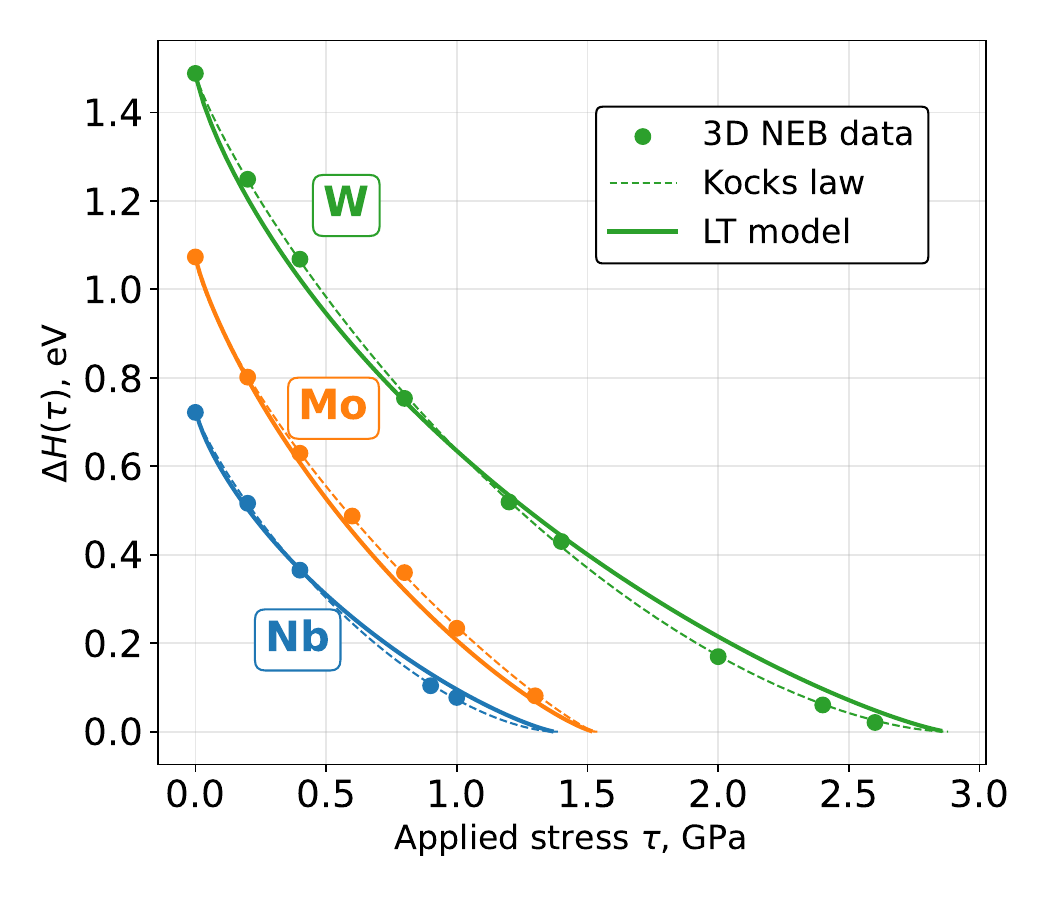}
    \caption{Activation enthalpy barriers for W (green), Mo (orange), and Nb (blue) as a function of applied stress. Symbols represent NEB data, solid lines show predictions from the line-tension (LT) model of Dorn and Rajnak~\cite{dorn1963nucleation}, and dashed lines correspond to fits based on Kocks' law.}
	\label{fig:enthalpy_barrier}
\end{figure}

Symbols in Fig.~\ref{fig:enthalpy_barrier} show the enthalpy barrier for screw dislocation motion as a function of applied stress in Nb, Mo, and W. At a given applied stress, the activation enthalpy follows the trend W~$>$~Mo~$>$~Nb.
Table~\ref{tab:kink_pair_energy} reports the kink-pair nucleation enthalpy at $T=0$~K and zero applied stress. ACE-2025 predictions are compared with values obtained from experimental fits~\cite{seeger2006slip,hollang1997flow,brunner2000comparison} and from a line-tension model parameterized using DFT Peierls potentials~\cite{dezerald2015first}. For Nb, ACE-2025 agrees well with the experimental value. For Mo and W, ACE-2025 is in better agreement with the DFT-parameterized line-tension model. The overall trend across the three RMs is also consistent with that reported previously for other bcc RMs~\cite{wen2000atomistic,yang2001kink,zotov2022entropy,maresca2018screw}. 

Table~\ref{tab:kink_pair_energy} also lists the Peierls stress at $T=0$~K obtained from direct MS simulations using the ACE-2025 potential, together with experimental measurements and DFT predictions. Note that the Peierls stress is derived from the Peierls potential according to $\tau_P=\mathrm{max}\left[(1/b)\, dE/dx\right]$, where $E$ is the Peierls potential and $b$ is the magnitude of the Burgers vector. ACE-2025 agrees well with the DFT results of Ref.~\cite{weinberger2013peierls} for Mo and W, but overestimates $\tau_P$ for Nb. Compared with experiment, ACE-2025 correctly predicts the ordering of the Peierls stress among the three elements, namely W~$>$~Mo~$>$~Nb. However, this discrepancy should not be viewed as a limitation specific to the present ACE-2025 model, but rather as a broader challenge in quantitatively predicting experimental Peierls stresses from atomistic calculations. This difficulty is associated with effects such as zero-point vibrations at low temperatures and non-glide stresses~\cite{proville2012quantum,dezerald2016plastic,vitek2004influence,groger2008multiscale}.

We next fit the simulation data using an exponential function. This phenomenological model, originally proposed by Kocks~\cite{kocks1975models}, has been widely applied to thermally activated dislocation motion and takes the general form
\begin{equation} \label{eq:kocks_law}
    {{\Delta }}H^\ast (\tau ) = {{\Delta }}H^\ast (0)\left[ {1 - \left( {\frac{\tau }{{\tau _P}}} \right)^p} \right]^q.
\end{equation}
Here, $\Delta H^\ast(0)$ and $\tau_P$ denote the activation enthalpy at zero stress and the Peierls stress, respectively, both of which can be obtained directly from atomistic simulations. The parameters $p$ and $q$ are determined by fitting. Using the ACE-2025 values of $\Delta H^\ast(0)$ and $\tau_P$ reported in Table~\ref{tab:kink_pair_energy}, we fit Eq.~\ref{eq:kocks_law} for Nb, Mo, and W. The fitted curves reproduce closely the simulation data, indicating that screw dislocation motion in all three RMs follows Kocks' law. The resulting fitting parameters $p$ and $q$ are listed in Table~\ref{tab:enthalpy_model_parameter}. For W, the fitted parameters are in very good agreement with those obtained previously using an EAM potential ($p=0.86$, $q=1.69$)~\cite{po2016phenomenological,marinica2013interatomic}.

\begin{table}[h!!!]
    \centering
		\renewcommand{\arraystretch}{1.2}
    \caption{Fitted parameters of Kocks' law ($p$ and $q$) and of the line-tension (LT) model ($\alpha$) for Nb, Mo, and W.}
    \begin{tabular}{c|c|ccc}
    \hline
    & Parameter & Nb & Mo & W  \\
    \hline
    \multirow{2}{*}{Kocks law}
     & $p$ & 0.796 & 0.742 & 0.854 \\
     & $q$ & 1.435 & 1.149 & 1.633\\
     \hline     
    LT model
    & $\alpha$ &  -1.114  & -0.203  & -0.631 \\
     \hline 
    \end{tabular}
    \label{tab:enthalpy_model_parameter}
\end{table}

Finally, we consider theoretical models for kink-pair nucleation in screw dislocations. To this end, we briefly review and apply the line-tension model of Dorn and Rajnak~\cite{dorn1963nucleation}. In this model, the dislocation is represented as a flexible line moving in the Peierls potential $V_P(x)$ under an applied stress. Changes in shape away from a straight dislocation are countered by the line energy $\Gamma_0$. The resulting energy functional describes the enthalpy of the dislocation as a function of its shape, and its saddle-point solution yields the critical kink-pair configuration and the corresponding enthalpy barrier, written as 
\begin{equation}
    {{\Delta }}H (\tau ) = 2{\int}_{x_0}^{x_c} \{ [\Gamma _0 + V_P(x)]^2 - [\tau b(x - x_0) + \Gamma _0 + V_P(x_0)]^2\} ^{\frac{1}{2}}{\rm d}x\ ,
\end{equation}
where $x_0$ represents the equilibrium configuration of the straight screw dislocation at the applied stress $\tau$, which can be obtained via
\begin{equation}
    \tau b = \frac{{\partial V_P(x)}}{{\partial x}}|_{x = x_0}.
\end{equation}
$x_c$ is the extremum position of the kink bulge in the critical configuration at the applied stress $\tau$, and it can be obtained by solving
 \begin{equation}
    \Delta H^\ast  (0) = 2\Gamma_0{\int}_{ - a/2}^{a/2} \sqrt {\left( {\frac{{\Gamma_0 + V_P(x)}}{{\Gamma_0}}} \right)^2 - 1} {\rm d}x\ ,
\end{equation}
where $\Delta H^\ast (0)$ is the kink-pair nucleation energy obtained by atomistic simulations. The Peierls potential is fitted to the following form
\begin{equation}
    V_P(x) = \frac{{{{\Delta }}E_P}}{2}\left[ {1 + \frac{\alpha }{4} + {{cos}}\left( {\frac{{2\pi x}}{a}} \right) - \frac{\alpha }{4}{{cos}}\left( {\frac{{4\pi x}}{a}} \right)} \right]~.
\end{equation}
$\alpha$ is chosen to match the Peierls stress from atomistic simulation results
\begin{equation}
    \tau _P = \frac{1}{b}\begin{array}{*{20}{c}} {} \\ {
    \rm max} \\ {\{ x \in [ - a/2,a/2]\} } \end{array}\left[ {\frac{{{d}}}{{{{d}}x}}V_P(x)} \right].
\end{equation}

Following this procedure, we apply the LT model for Nb, Mo, and W, and the resulting parameters are listed in Table~\ref{tab:enthalpy_model_parameter}. The fitting uses the values of $\Delta H^\ast(0)$ and $\tau_P$ predicted by the ACE-2025 potential (Table~\ref{tab:kink_pair_energy}). As shown by the solid lines in Fig.~\ref{fig:enthalpy_barrier}, the LT model reproduces both the NEB data and the fitted Kocks law very well. This result indicates that the enthalpy barrier as a function of applied stress can be described accurately using only the zero-stress activation enthalpy and the Peierls stress, thereby significantly reducing the computational cost \cite{rodney2009stress,allera2025activation}.

\section{Discussion}\label{sec:discussion}

\subsection{Peierls potential in bcc RMs}

Accurately predicting screw dislocation behavior at the atomic scale has remained a longstanding challenge in bcc RMs. Classical IAPs often predict degenerate core structures and double-humped Peierls potentials~\cite{proville2012quantum,Gordon_2011,starikov2024angular}. More recently, ML-IAPs have been developed and benchmarked against screw dislocation properties, especially for W~\cite{szlachta2014accuracy,goryaeva2021efficient}. However, comparable models for other bcc RMs remain scarce, and comprehensive validation across multiple RMs is still missing. As shown in Fig.~\ref{fig:gap_predictions}, even recent ML-IAPs cannot consistently reproduce DFT Peierls barriers across the bcc RMs considered here~\cite{byggmastar2026nine}. In this work, we developed a family of ACE-2025 potentials for bcc RMs and validated them against elastic and surface properties, as well as screw dislocation behavior. In the remainder of this subsection, we discuss the DFT database design required to reproduce the Peierls potential and its implications for predicting the more complex energetics of screw dislocations in bcc RMs.

To obtain accurate Peierls potentials in V, Nb, and Ta, we gradually enriched the DFT database by adding stacking fault configurations, equations of state, and screw dislocation structures. For V and Ta, including stacking faults and equations of state is sufficient for ACE-2025 to reproduce the Peierls potential with near-DFT accuracy. For Nb, however, additional screw dislocation configurations are required to recover the DFT barrier. This observation suggests that the transferability of the potential to screw dislocation properties depends not only on the configuration types represented in the training database but also on the element specific energy landscape.

Earlier work on W based on GAP showed that including vacancy-containing $\gamma$ surfaces can worsen the description of the Peierls barrier, whereas adding screw dislocation quadrupole cells restores quantitative agreement with DFT~\cite{szlachta2014accuracy}. That study further demonstrated that a database composed only of screw dislocation configurations can quantitatively reproduce the screw dislocation structure, dislocation-vacancy binding energies, and the Peierls barrier. By contrast, our results indicate that, within the ACE framework, a database containing only $\gamma$ surfaces is already sufficient to recover the screw dislocation energy landscape with near-DFT accuracy for both W and Mo. This difference may reflect the more complete representation of local atomic environments provided by the ACE descriptor~\cite{dusson2022atomic}. This observation suggests that the predictive capability of ML-IAPs for dislocations depends not only on the training data but also on the representation of the atomic environments. Clarifying this interplay would require another systematic investigation of database design and model representation. 

\begin{figure}[h!!!]
	\centering
	\includegraphics[width=0.8\linewidth]{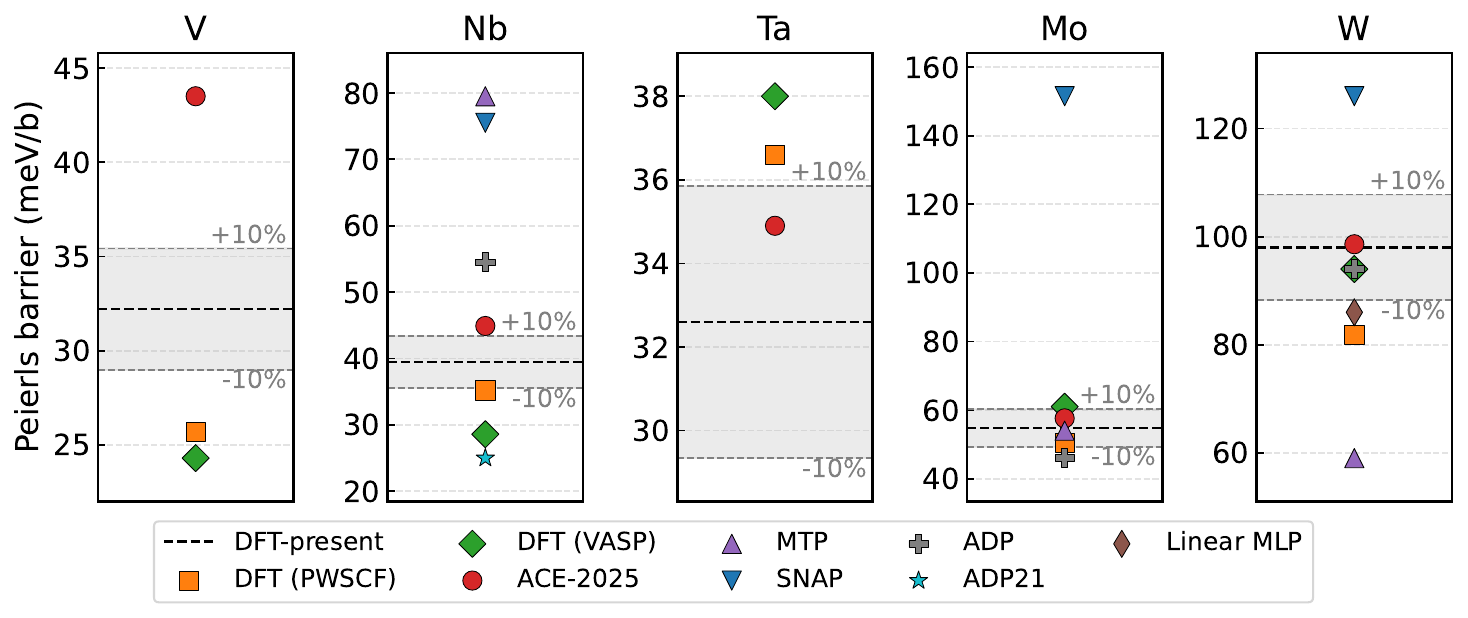}
    \caption{Comparison of Peierls barriers for V, Nb, Ta, Mo, and W predicted by ACE-2025 and selected interatomic potentials. The black dashed line denotes the DFT prediction from present calculations, and the shaded region indicates the corresponding $\pm 10\%$ deviation. 
    Literature DFT values computed using PWSCF~\cite{dezerald2014ab} and VASP~\cite{weinberger2013peierls} are included for comparison. 
    MTP~\cite{yin2021atomistic}, SNAP~\cite{li2020complex}, and ADP~\cite{starikov2024angular} results are taken from Ref.~\cite{starikov2024angular}; the ADP21 prediction is taken from Ref.~\cite{starikov2021optimized}; and the Linear MLP prediction for W is taken from Ref.~\cite{leveau2025segregation}. Note that the Peierls barriers predicted by ADP exhibit a "doubly-humped" feature with a local minimum corresponding to an intermediate metastable configuration.}
	\label{fig:PB_comparison}
\end{figure}

Fig.~\ref{fig:PB_comparison} compares the Peierls barriers computed from DFT and different IAPs. The DFT results consistently show that the Peierls barrier increases from group 5 to group 6 RMs, with W exhibiting the highest barrier~\cite{dezerald2014ab,weinberger2013peierls}. The present DFT calculations are close to previous DFT results, although slight deviations remain for W. ACE-2025 reproduces the overall trend across the five RMs and yields values that are broadly comparable to the DFT reference data. In contrast, larger discrepancies are observed for several previously reported ML and classical IAPs. In particular, SNAP strongly overestimates the Peierls barrier for Mo and W, whereas ADP predicts a doubly humped barrier for Nb, Mo, and W, indicating the presence of an intermediate metastable configuration~\cite{starikov2024angular}. These comparisons show that, among the available IAPs for bcc RMs, only ACE-2025 achieves quantitative agreement with DFT across multiple elements. 

Although ACE-2025 accurately reproduces the Peierls barrier along the minimum energy path of screw dislocations, a faithful description of the more complex two-dimensional energy landscape, including the energetics of the hard and split cores, remains challenging for V and Ta. This limitation is reflected in the less accurate prediction of the deviation angle along the core trajectory in V and Ta (Table~\ref{tab:core_traj}), as well as in the difficulty of reproducing the correct energy hierarchy of the hard and split cores in Ta. These results indicate that ACE-2025 still requires further refinement for V and Ta, while also demonstrating that the Peierls barrier alone is an incomplete metric for validating IAPs in bcc RMs. For example, ADP21 reproduces a single-hump Peierls potential, but nevertheless fails to predict screw-dislocation glide on the $\{110\}$ plane at $T=240$~K~\cite{starikov2021optimized,starikov2025dislocation}.  A reliable potential should reproduce the correct energetic hierarchy of the three core structures (easy, hard, and split) before being applied in large scale dislocation simulations. Ideally, validating the dislocation properties of elementary bcc RMs should be regarded as a necessary step before extending the IAPs to multi-element systems, such as refractory high-entropy alloys.

\subsection{Scaling strategy for constructing DFT database}

The differences in the accuracy of different elements motivate a closer examination of the strategy for constructing the underlying DFT database. Across all validation tests, we found a consistent trend: ACE-2025 achieves near-DFT accuracy for the group 6 elements, Mo and W, whereas its performance becomes less accurate for the group 5 elements (V, Nb, Ta). For Mo and W, ACE-2025 reproduces bulk properties such as elastic constants, surface energies, and vacancy formation energies, while also capturing more subtle features, including the Peierls barrier and the full screw dislocation energy landscape.

The strong dependence on the chemical group most likely originates from the construction of the underlying DFT database. The original training set was designed for W and then transferred to the other elements by a simple rescaling of the lattice parameter. For V, Nb, Ta, and Mo, DFT energies and forces were generated by performing only self consistent electronic structure calculations at these rescaled geometries, without relaxing either the ionic positions or the simulation cell. As a consequence, a substantial fraction of the database for these elements may correspond to nonequilibrium configurations and may not adequately sample the relevant local minima and transition states on the true potential energy surface. This problem is particularly important for elements outside the group 6 column, where bonding characteristics and equilibrium structures differ significantly from those of W, even though all elements share the same bcc crystal structure. The present results expose intrinsic limitations of simple lattice-parameter rescaling strategies for building DFT databases. Rescaling a database constructed for W works well for Mo, which belongs to the same chemical group, but fails to provide equally accurate models for V, Nb and Ta.

These findings highlight an important limitation of constructing multi-element DFT databases by straightforward lattice parameter rescaling, even when the materials have identical crystal symmetry~\cite{byggmastar2021modeling,byggmastar2022simple,byggmastar2026nine}. While such a strategy is attractive from a computational efficiency perspective, it can introduce systematic biases into the training data and, in turn, into the learned interatomic potentials, especially for complex defect configurations such as dislocation cores. Our results emphasize the need for element-specific, fully relaxed DFT configurations or active learning strategies when high fidelity is required to capture the detailed atomistic energy landscape of complex defects.

\section{Conclusion}\label{sec:conclusion}
In this work, we developed ACE-2025 potentials for five bcc RMs and benchmarked a broad range of screw dislocation properties. The main conclusions are as follows:

\begin{itemize}
    \item None of the existing ML-IAPs can provide a consistently accurate description of screw dislocation Peierls barriers across V, Nb, Ta, Mo, and W. By extending a publicly available DFT database and training within the ACE framework, we obtained ACE-2025 potentials that significantly improve the description of screw dislocation properties, achieving near-DFT accuracy for all RMs.
    
    \item The transferability of ML-IAPs to screw dislocation properties depends sensitively on both the composition of the training database and the specific element considered. In particular, the present results show that the Peierls barrier alone is not a sufficient validation metric. A reliable validation should also include the two-dimensional energy landscape of the screw dislocation. 
    
    \item For Nb, Mo, and W, the ACE-2025 potentials enable reliable calculations of kink-pair activation enthalpies under applied stress. The atomistic results are well described by both Kocks' law and a LT model, providing an efficient route to connect atomistic screw dislocation energetics with dislocation mobility models in bcc RMs.

    \item The present results also reveal an important limitation of the lattice rescaling strategy used to construct the original multi-element DFT database. While this approach works reasonably well for Mo and W, it is less effective for V, Nb, and Ta, indicating that more element-specific relaxed configurations are needed when high-fidelity descriptions of complex defect energetics are required.
\end{itemize}

\newpage

\appendix
\markright{Appendix}
\renewcommand{\thefigure}{A.\arabic{figure}} 
\renewcommand{\thetable}{A.\arabic{table}} 
\setcounter{table}{0}
\setcounter{figure}{0}
\captionsetup{listof=false}

\section{Structures in the DFT database}\label{append:dft_data}
\begin{table}[h!!!]
    \centering
    \caption{Summary of the enriched DFT database for each RM.}
    \begin{tabular}{c|c|c|c}
        \hline
        Element & Configuration & Number of structures & Number of LAEs \\
        \hline
        \multirow{7}{*}{V}
           & primitve  & 2963 & 2963  \\
           & traction-separation & 276 & 6624  \\
		   & Stacking fault \{110\} & 314 & 6280 \\ 
		   & Stacking falut \{112\} & 380 & 11400 \\ 
           & EOS of fcc & 80 & 80 \\
           & EOS of bcc & 80 & 80 \\
           & EOS of hcp & 180 & 360 \\
        \hline
        \multirow{8}{*}{Nb}
          &  primitve  & 2963 & 2963  \\
		  &  traction-separation & 276  & 6624 \\
		  & Stacking fault \{110\} & 826 & 16520 \\ 
		  & Stacking falut \{112\} & 450 & 13500 \\ 
	      & EOS of fcc & 80 & 80 \\
		  & EOS of bcc & 80 & 80 \\
	      & EOS of hcp & 180 & 360 \\
          & Screw dislocation & 29 & 3915 \\
        \hline
        \multirow{5}{*}{Ta}
          & primitve  & 2963 & 2963  \\
		  & traction-separation & 276  & 6624\\
		  & EOS of fcc & 30 & 30 \\
		  & EOS of bcc & 30 & 30 \\
		  & EOS of hcp & 30 & 60 \\
        \hline
        \multirow{2}{*}{Mo}
          & primitve  & 2963 & 2963  \\
		  & traction-separation & 276  & 6624\\
        \hline
        \multirow{2}{*}{W}
        &  primitve  & 2963 & 2963  \\
		&  traction-separation & 276 & 6624  \\
        \hline
    \end{tabular}
    \label{tab:dft_database_summary}
\end{table}

\begin{table}[h]
	\centering
	\small
	\caption{Configuration types included in the original GAP-2020 DFT database for W. The same configuration types are used for the other RMs~\cite{byggmastar2020gaussian}.}
	\begin{tabular}{l|c|c}
		\hline
		Configuration & Number of structures & Number of LAEs \\
		\hline
		A15            & 100  & 800   \\
		bcc distorted  & 500  & 1000  \\
		C15            & 100  & 600   \\
		di-SIA         & 15   & 2350  \\
		di-vacancy     & 10   & 1180  \\
		diamond        & 100  & 200   \\
		dimer          & 13   & 26    \\
		fcc            & 100  & 100   \\
		gamma surface  & 178  & 2136  \\
		hcp            & 100  & 200   \\
		isolated atom  & 1    & 1     \\
		liquid         & 45   & 5760  \\
		phonon         & 20   & 1080  \\
		sc             & 100  & 100   \\
		short range    & 90   & 4860  \\
		SIA            & 32   & 3872  \\
		slice sample   & 1996 & 1996  \\
		surf liquid    & 24   & 3264  \\
		surface (100)  & 45   & 540   \\
		surface (110)  & 45   & 540   \\
		surface (111)  & 43   & 516   \\
		surface (112)  & 45   & 540   \\
		tri-vacancy    & 15   & 1755  \\
		vacancy        & 210  & 11130 \\
		\hline
		Total          & 3927 & 44546 \\
		\hline
	\end{tabular}
	\label{tab:gap2020_database_configurations}
\end{table}

\section{Principal component analysis of the DFT database}\label{append:pca_analysis}
\begin{figure}[h!!!]
	\centering
	\includegraphics[width=0.4\linewidth]{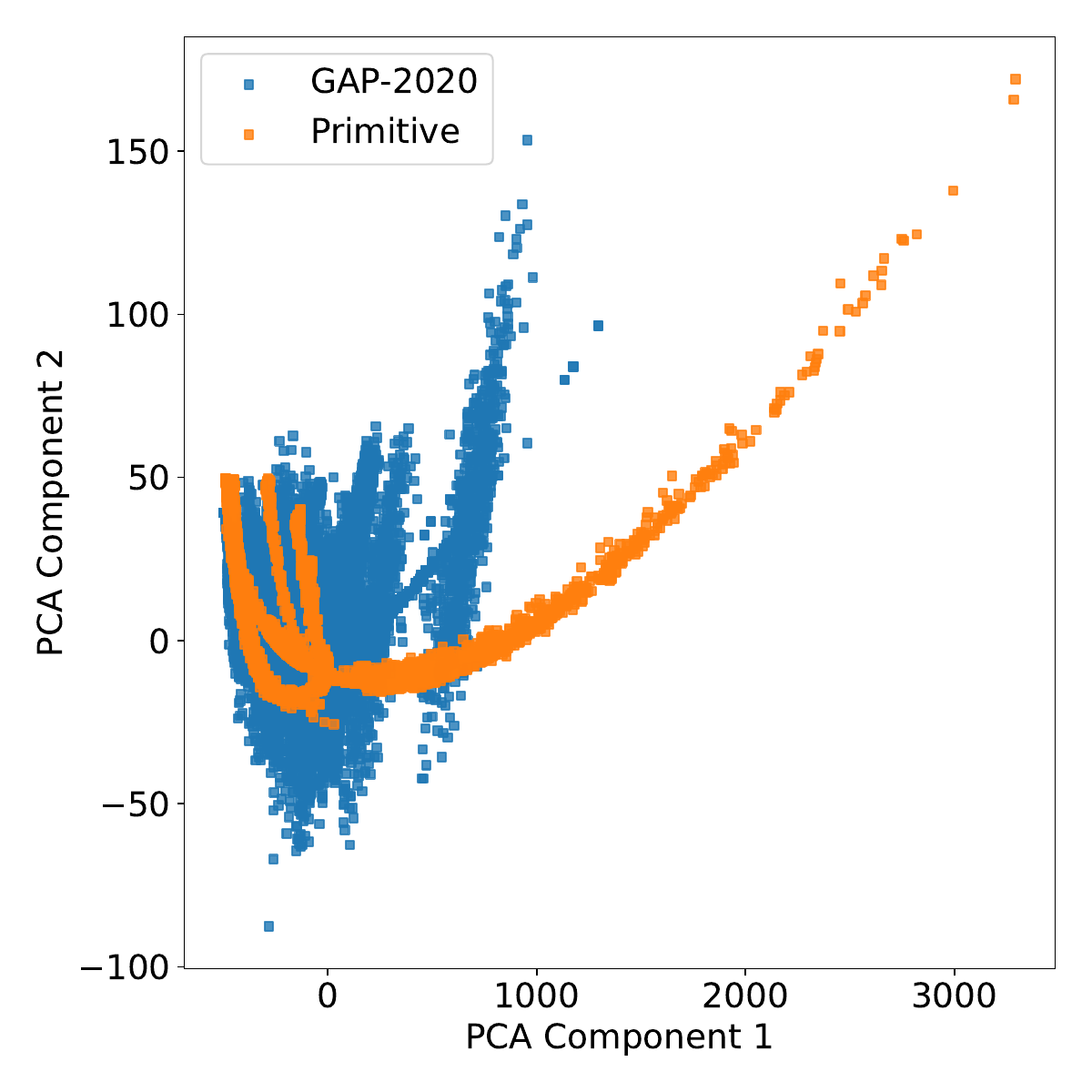}
	\caption{Principal component analysis projection of the tungsten database in the ACE B-basis descriptor space. }
	\label{fig:pca_db}
\end{figure}

Fig.~\ref{fig:pca_db} compares the extended primitive cell configurations with the original GAP training database, showing how the newly generated configurations are distributed relative to the reference dataset in descriptor space. The primitive cells were generated by rescaling primitive cell configurations previously constructed for iron~\cite{zhang2023atomistic}. These cells were subjected to large strains in the six independent strain components, $\varepsilon_{ij}$, ranging from $-0.22$ to $0.3$, with the strain space sampled using a Sobol sequence~\cite{sobol1967distribution}.

\section{Extrapolation grade  $\gamma$}

\begin{figure}[h!!!]
	\centering
	\includegraphics[width=\linewidth]{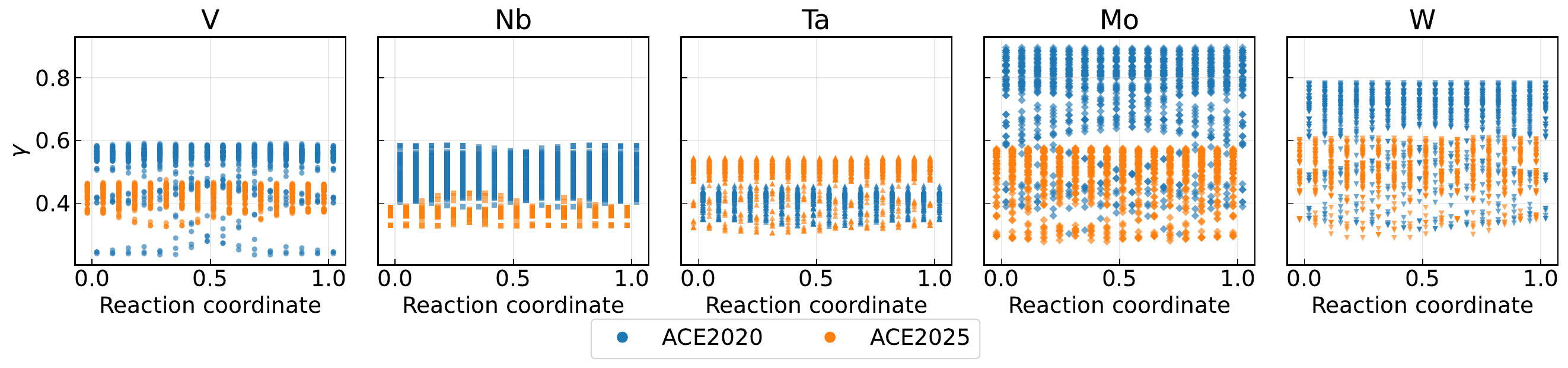}
	\caption{Comparison of the extrapolation index ($\gamma$) distributions along the NEB reaction coordinate predicted by ACE2020 and ACE2025 for the five RMs. Each panel corresponds to one element, and the scattered points represent the $\gamma$ values predicted from different replicas at each reaction coordinate. The two colors distinguish the ACE2020 and ACE2025 predictions.}
	\label{fig:gamma_neb}
\end{figure}

\begin{figure}[h!!!]
	\centering
	\includegraphics[width=0.5\linewidth]{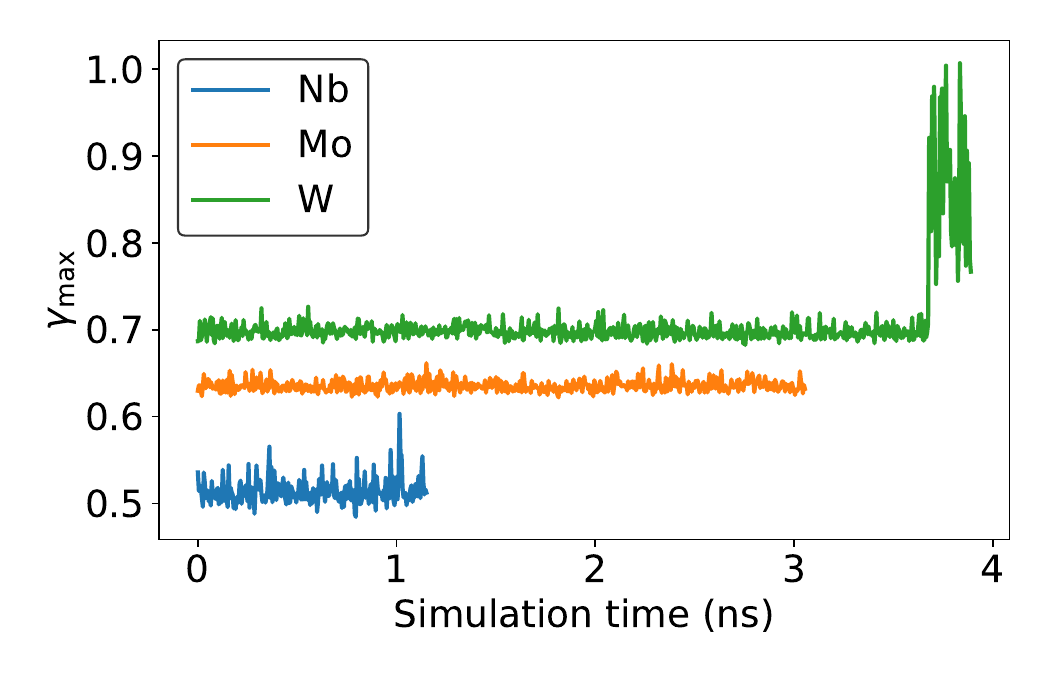}
	\caption{Maximum extrapolation grade, $\gamma_{\rm max}$, as a function of simulation time for Nb, Mo, and W during shear loading at $T=200~\rm{K}$, using ACE2025. The $\gamma_{\rm max}$ is the maximum extrapolation grade among all atoms at each timestep.}
	\label{fig:gamma_finiteT}
\end{figure}

\FloatBarrier
\noindent\textbf{Data Availability}\\
LAMMPS and VASP inputs, DFT databases, and ACE potential files are available on our Github page \url{https://github.com/leiapple/ACE-BCC-TMs}. Python scripts for the line-tension model and screw dislocation analysis, including Peierls potential and trajectories, are available on the same Github page.  \\

\noindent\textbf{Acknowledgments}\\
This work used the Dutch national electronic infrastructure with the support of the SURF Cooperative using grant no. EINF-9170. The authors thank the Center for Information Technology of the University of Groningen for their support and for providing access to the Hábrók high performance computing cluster. The discussion with Fei Shuang from Delft University and Fan-shun Meng from the University of Osaka are highly appreciated. \\

\noindent\textbf{Author contributions}\\
L.Z. and F.M. designed the research together. L.Z. performed all the DFT, MS, and MD calculations. All authors analyzed the data, discussed the results, wrote the manuscript together, and contributed to the discussions and revisions of the paper.  \\

\noindent\textbf{Competing interests}\\
The authors declare no competing interests.\\

\medskip
\printbibliography

@article{maresca2018screw,
	title={Screw dislocation structure and mobility in body centered cubic Fe predicted by a Gaussian Approximation Potential},
	author={Maresca, Francesco and Dragoni, Daniele and Cs{\'a}nyi, G{\'a}bor and Marzari, Nicola and Curtin, William A.},
	journal={npj Computational Materials},
	volume={4},
	pages={69},
	year={2018},
	publisher={Nature Publishing Group UK London}
}

@article{zhang2023atomistic,
	title={Atomistic fracture in bcc iron revealed by active learning of Gaussian approximation potential},
	author={Zhang, Lei and Cs{\'a}nyi, G{\'a}bor and Van Der Giessen, Erik and Maresca, Francesco},
	journal={npj Computational Materials},
	volume={9},
	number={1},
	pages={217},
	year={2023},
	publisher={Nature Publishing Group UK London}
}

@article{goryaeva2021efficient,
	title={Efficient and transferable machine learning potentials for the simulation of crystal defects in bcc Fe and W},
	author={Goryaeva, Alexandra M and D{\'e}r{\`e}s, Julien and Lapointe, Clovis and Grigorev, Petr and Swinburne, Thomas D and Kermode, James R and Ventelon, Lisa and Baima, Jacopo and Marinica, Mihai-Cosmin},
	journal={Physical Review Materials},
	volume={5},
	number={10},
	pages={103803},
	year={2021},
	publisher={APS}
}

@article{byggmastar2020gaussian,
	title={Gaussian approximation potentials for body-centered-cubic transition metals},
	author={Byggm{\"a}star, Jesper and Nordlund, Kai and Djurabekova, Flyura},
	journal={Physical Review Materials},
	volume={4},
	number={9},
	pages={093802},
	year={2020},
	publisher={APS}
}

@article{zhang2024efficiency,
	title={Efficiency, accuracy, and transferability of machine learning potentials: Application to dislocations and cracks in iron},
	author={Zhang, Lei and Cs{\'a}nyi, G{\'a}bor and van der Giessen, Erik and Maresca, Francesco},
	journal={Acta Materialia},
	volume={270},
	pages={119788},
	year={2024},
	publisher={Elsevier}
}

@article{wang2022classical,
	title={Classical and machine learning interatomic potentials for BCC vanadium},
	author={Wang, Rui and Ma, Xiaoxiao and Zhang, Linfeng and Wang, Han and Srolovitz, David J and Wen, Tongqi and Wu, Zhaoxuan},
	journal={Physical Review Materials},
	volume={6},
	number={11},
	pages={113603},
	year={2022},
	publisher={APS}
}

@article{kwon2023accurate,
	title={Accurate description of hydrogen diffusivity in bcc metals using machine-learning moment tensor potentials and path-integral methods},
	author={Kwon, Hyukjoon and Shiga, Motoyuki and Kimizuka, Hajime and Oda, Takuji},
	journal={Acta Materialia},
	volume={247},
	pages={118739},
	year={2023},
	publisher={Elsevier}
}

@article{ito2024machine,
	title={Machine learning interatomic potential with DFT accuracy for general grain boundaries in $\alpha$-Fe},
	author={Ito, Kazuma and Yokoi, Tatsuya and Hyodo, Katsutoshi and Mori, Hideki},
	journal={npj Computational Materials},
	volume={10},
	number={1},
	pages={255},
	year={2024},
	publisher={Nature Publishing Group UK London}
}

@article{nikoulis2021machine,
	title={Machine-learning interatomic potential for W--Mo alloys},
	author={Nikoulis, Giorgos and Byggm{\"a}star, Jesper and Kioseoglou, Joseph and Nordlund, Kai and Djurabekova, Flyura},
	journal={Journal of Physics: Condensed Matter},
	volume={33},
	number={31},
	pages={315403},
	year={2021},
	publisher={IOP Publishing}
}

@article{yin2021atomistic,
	title={Atomistic simulations of dislocation mobility in refractory high-entropy alloys and the effect of chemical short-range order},
	author={Yin, Sheng and Zuo, Yunxing and Abu-Odeh, Anas and Zheng, Hui and Li, Xiang-Guo and Ding, Jun and Ong, Shyue Ping and Asta, Mark and Ritchie, Robert O},
	journal={Nature communications},
	volume={12},
	number={1},
	pages={4873},
	year={2021},
	publisher={Nature Publishing Group UK London}
}

@article{lyu2023effects,
	title={Effects of chemical randomness on strength contributors and dislocation behaviors in a bcc multiprincipal element alloy},
	author={Lyu, Shuang and Li, Wei and Xia, Yuanhang and Chen, Yue and Ngan, Alfonso HW},
	journal={Physical Review Materials},
	volume={7},
	number={7},
	pages={073602},
	year={2023},
	publisher={APS}
}

@article{dezerald2014ab,
	title={Ab initio modeling of the two-dimensional energy landscape of screw dislocations in bcc transition metals},
	author={Dezerald, L and Ventelon, Lisa and Clouet, E and Denoual, C and Rodney, David and Willaime, F},
	journal={Physical Review B},
	volume={89},
	number={2},
	pages={024104},
	year={2014},
	publisher={APS}
}

@article{dezerald2016plastic,
  title={Plastic anisotropy and dislocation trajectory in BCC metals},
  author={Dezerald, Lucile and Rodney, David and Clouet, Emmanuel and Ventelon, Lisa and Willaime, Francois},
  journal={Nature communications},
  volume={7},
  number={1},
  pages={11695},
  year={2016},
  publisher={Nature Publishing Group UK London}
}

@article{lysogorskiy2021performant,
	title={Performant implementation of the atomic cluster expansion (PACE) and application to copper and silicon},
	author={Lysogorskiy, Yury and Oord, Cas van der and Bochkarev, Anton and Menon, Sarath and Rinaldi, Matteo and Hammerschmidt, Thomas and Mrovec, Matous and Thompson, Aidan and Cs{\'a}nyi, G{\'a}bor and Ortner, Christoph and others},
	journal={npj Computational Materials},
	volume={7},
	number={1},
	pages={97},
	year={2021},
	publisher={Nature Publishing Group UK London}
}

@article{bochkarev2022efficient,
  title={Efficient parametrization of the atomic cluster expansion},
  author={Bochkarev, Anton and Lysogorskiy, Yury and Menon, Sarath and Qamar, Minaam and Mrovec, Matous and Drautz, Ralf},
  journal={Physical Review Materials},
  volume={6},
  number={1},
  pages={013804},
  year={2022},
  publisher={APS}
}

@article{pan2024atomic,
  title={Atomic cluster expansion interatomic potential for defects and thermodynamics of Cu--W system},
  author={Pan, Jiahao and Cheng, Huiqun and Yan, Gaosheng and Zhang, Lei and Yu, Wenshan and Shen, Shengping},
  journal={Journal of Applied Physics},
  volume={136},
  number={15},
  year={2024},
  publisher={AIP Publishing}
}

@article{perdew1996generalized,
  title={Generalized gradient approximation made simple},
  author={Perdew, John P and Burke, Kieron and Ernzerhof, Matthias},
  journal={Physical review letters},
  volume={77},
  number={18},
  pages={3865},
  year={1996},
  publisher={APS}
}

@article{kresse1993ab,
  title={Ab initio molecular dynamics for liquid metals},
  author={Kresse, Georg and Hafner, J{\"u}rgen},
  journal={Physical Review B},
  volume={47},
  number={1},
  pages={558},
  year={1993},
  publisher={APS}
}

@article{kresse1994ab,
  title={Ab initio molecular-dynamics simulation of the liquid-metal--amorphous-semiconductor transition in germanium},
  author={Kresse, Georg and Hafner, J{\"u}rgen},
  journal={Physical Review B},
  volume={49},
  number={20},
  pages={14251},
  year={1994},
  publisher={APS}
}

@article{kresse1996efficiency,
  title={Efficiency of ab-initio total energy calculations for metals and semiconductors using a plane-wave basis set},
  author={Kresse, Georg and Furthm{\"u}ller, J{\"u}rgen},
  journal={Computational Materials Science},
  volume={6},
  number={1},
  pages={15--50},
  year={1996},
  publisher={Elsevier}
}

@article{kresse1996efficient,
  title={Efficient iterative schemes for ab initio total-energy calculations using a plane-wave basis set},
  author={Kresse, Georg and Furthm{\"u}ller, J{\"u}rgen},
  journal={Physical Review B},
  volume={54},
  number={16},
  pages={11169},
  year={1996},
  publisher={APS}
}

@article{thomas2024hyperparameter,
  title={Hyperparameter optimization for atomic cluster expansion potentials},
  author={Thomas du Toit, Daniel F and Zhou, Yuxing and Deringer, Volker L},
  journal={Journal of Chemical Theory and Computation},
  volume={20},
  number={22},
  pages={10103--10113},
  year={2024},
  publisher={ACS Publications}
}

@article{henkelman2000climbing,
  title={A climbing image nudged elastic band method for finding saddle points and minimum energy paths},
  author={Henkelman, Graeme and Uberuaga, Blas P and J{\'o}nsson, Hannes},
  journal={The Journal of chemical physics},
  volume={113},
  number={22},
  pages={9901--9904},
  year={2000},
  publisher={American Institute of Physics}
}

@article{li2004core,
  title={Core energy and Peierls stress of a screw dislocation in bcc molybdenum: A periodic-cell tight-binding study},
  author={Li, Ju and Wang, Cai-Zhuang and Chang, Jin-Peng and Cai, Wei and Bulatov, Vasily V and Ho, Kai-Ming and Yip, Sidney},
  journal={Physical Review B},
  volume={70},
  number={10},
  pages={104113},
  year={2004},
  publisher={APS}
}

@article{ventelon2013ab,
  title={Ab initio investigation of the Peierls potential of screw dislocations in bcc Fe and W},
  author={Ventelon, Lisa and Willaime, Fran{\c{c}}ois and Clouet, Emmanuel and Rodney, David},
  journal={Acta Materialia},
  volume={61},
  number={11},
  pages={3973--3985},
  year={2013},
  publisher={Elsevier}
}

@article{stroh1958dislocations,
  title={Dislocations and cracks in anisotropic elasticity},
  author={Stroh, AN},
  journal={Philosophical Magazine},
  volume={3},
  number={30},
  pages={625--646},
  year={1958},
  publisher={Taylor \& Francis}
}

@misc{atomman,
  url = {https://www.ctcms.nist.gov/potentials/atomman/}
}

@article{cai2003periodic,
  title={Periodic image effects in dislocation modelling},
  author={Cai, Wei and Bulatob, Vasily V and Chang, Jinpeng and Li, Ju and Yip, Sidney},
  journal={Philosophical Magazine},
  volume={83},
  number={5},
  pages={539--567},
  year={2003},
  publisher={Taylor \& Francis}
}

@article{erhard2024modelling,
  title={Modelling atomic and nanoscale structure in the silicon--oxygen system through active machine learning},
  author={Erhard, Linus C and Rohrer, Jochen and Albe, Karsten and Deringer, Volker L},
  journal={Nature Communications},
  volume={15},
  number={1},
  pages={1927},
  year={2024},
  publisher={Nature Publishing Group UK London}
}

@article{drautz2019atomic,
  title={Atomic cluster expansion for accurate and transferable interatomic potentials},
  author={Drautz, Ralf},
  journal={Physical Review B},
  volume={99},
  number={1},
  pages={014104},
  year={2019},
  publisher={APS}
}

@article{larsen2017atomic,
  title={The atomic simulation environment—a Python library for working with atoms},
  author={Larsen, Ask Hjorth and Mortensen, Jens J{\o}rgen and Blomqvist, Jakob and Castelli, Ivano E and Christensen, Rune and Du{\l}ak, Marcin and Friis, Jesper and Groves, Michael N and Hammer, Bj{\o}rk and Hargus, Cory and others},
  journal={Journal of Physics: Condensed Matter},
  volume={29},
  number={27},
  pages={273002},
  year={2017},
  publisher={IOP Publishing}
}

@article{bitzek2006structural,
  title={Structural relaxation made simple},
  author={Bitzek, Erik and Koskinen, Pekka and G{\"a}hler, Franz and Moseler, Michael and Gumbsch, Peter},
  journal={Physical Review Letters},
  volume={97},
  number={17},
  pages={170201},
  year={2006},
  publisher={APS}
}

@article{thompson2022lammps,
  title={LAMMPS-a flexible simulation tool for particle-based materials modeling at the atomic, meso, and continuum scales},
  author={Thompson, Aidan P and Aktulga, H Metin and Berger, Richard and Bolintineanu, Dan S and Brown, W Michael and Crozier, Paul S and In't Veld, Pieter J and Kohlmeyer, Axel and Moore, Stan G and Nguyen, Trung Dac and others},
  journal={Computer Physics Communications},
  volume={271},
  pages={108171},
  year={2022},
  publisher={Elsevier}
}

@article{stukowski2009visualization,
  title={Visualization and analysis of atomistic simulation data with OVITO--the Open Visualization Tool},
  author={Stukowski, Alexander},
  journal={Modelling and Simulation in Materials Science and Engineering},
  volume={18},
  number={1},
  pages={015012},
  year={2009},
  publisher={IOP Publishing}
}

@article{zhang2023unstable,
  title={Unstable stacking fault energy and peierls stress for evaluating slip system competition in body-centered cubic metals},
  author={Zhang, Xue-Chun and Cao, Shuo and Zhang, Lian-Ji and Yang, Rui and Hu, Qing-Miao},
  journal={Journal of Materials Research and Technology},
  volume={22},
  pages={3413--3422},
  year={2023},
  publisher={Elsevier}
}

@article{szlachta2014accuracy,
  title={Accuracy and transferability of Gaussian approximation potential models for tungsten},
  author={Szlachta, Wojciech J and Bart{\'o}k, Albert P and Cs{\'a}nyi, G{\'a}bor},
  journal={Physical Review B},
  volume={90},
  number={10},
  pages={104108},
  year={2014},
  publisher={APS}
}

@article{grigorev2024matscipy,
  title={matscipy: materials science at the atomic scale with Python},
  author={Grigorev, Petr and Fr{\'e}rot, Lucas and Birks, Fraser and Gola, Adrien and Golebiowski, Jacek and Grie{\ss}er, Jan and H{\"o}rmann, Johannes L and Klemenz, Andreas and Moras, Gianpietro and N{\"o}hring, Wolfram G and others},
  journal={The Journal of Open Source Software (JOSS)},
  volume={9},
  number={93},
  year={2024}
}

@article{shuang2025modeling,
  title={Modeling extensive defects in metals through classical potential-guided sampling and automated configuration reconstruction},
  author={Shuang, Fei and Liu, Kai and Ji, Yucheng and Gao, Wei and Laurenti, Luca and Dey, Poulumi},
  journal={npj Computational Materials},
  volume={11},
  number={1},
  pages={118},
  year={2025},
  publisher={Nature Publishing Group UK London}
}

@article{dusson2022atomic,
  title={Atomic cluster expansion: Completeness, efficiency and stability},
  author={Dusson, Genevieve and Bachmayr, Markus and Cs{\'a}nyi, G{\'a}bor and Drautz, Ralf and Etter, Simon and van Der Oord, Cas and Ortner, Christoph},
  journal={Journal of Computational Physics},
  volume={454},
  pages={110946},
  year={2022},
  publisher={Elsevier}
}

@article{shuang2025universal,
  title={Universal machine learning interatomic potentials poised to supplant DFT in modeling general defects in metals and random alloys},
  author={Shuang, Fei and Wei, Zixiong and Liu, Kai and Gao, Wei and Dey, Poulumi},
  journal={Machine Learning: Science and Technology},
  volume={6},
  number={3},
  pages={030501},
  year={2025},
  publisher={IOP Publishing}
}

@article{byggmastar2026nine,
  title={Nine-element machine-learned interatomic potentials for multiphase refractory alloys},
  author={Byggm{\"a}star, Jesper and Lopes, Tiago and Fan, Zheyong and Ala-Nissila, Tapio},
  journal={arXiv preprint arXiv:2603.04147},
  year={2026}
}

@article{byggmastar2022simple,
  title={Simple machine-learned interatomic potentials for complex alloys},
  author={Byggm{\"a}star, Jesper and Nordlund, Kai and Djurabekova, Flyura},
  journal={Physical Review Materials},
  volume={6},
  number={8},
  pages={083801},
  year={2022},
  publisher={APS}
}

@article{wang2025ductility,
  title={Ductility mechanisms in complex concentrated refractory alloys from atomistic fracture simulations},
  author={Wang, Wenqing and Kumar, Punit and Cook, David H and Walsh, Flynn and Zhang, Buyu and Borges, Pedro PPO and Farkas, Diana and Ritchie, Robert O and Asta, Mark},
  journal={npj Computational Materials},
  volume={11},
  number={1},
  pages={330},
  year={2025},
  publisher={Nature Publishing Group UK London}
}

@article{hollang1997flow,
  title={The Flow Stress of Ultra-High-Purity Molybdenum Single Crystals},
  author={Hollang, L and Hommel, M and Seeger, A},
  journal={Physica status solidi (a)},
  volume={160},
  number={2},
  pages={329--354},
  year={1997},
  publisher={Wiley Online Library}
}

@article{seeger2006slip,
  title={Slip planes and kink properties of screw dislocations in high-purity niobium},
  author={Seeger, Alfred and Holzwarth, Uwe},
  journal={Philosophical Magazine},
  volume={86},
  number={25-26},
  pages={3861--3892},
  year={2006},
  publisher={Taylor \& Francis}
}

@article{brunner2000comparison,
  title={Comparison of flow-stress measurements on high-purity tungsten single crystals with the kink-pair theory},
  author={Brunner, Dieter},
  journal={Materials Transactions, JIM},
  volume={41},
  number={1},
  pages={152--160},
  year={2000},
  publisher={The Japan Institute of Metals}
}

@article{weinberger2013peierls,
  title={Peierls potential of screw dislocations in bcc transition metals: Predictions from density functional theory},
  author={Weinberger, Christopher R and Tucker, Garritt J and Foiles, Stephen M},
  journal={Physical Review B},
  volume={87},
  number={5},
  pages={054114},
  year={2013},
  publisher={APS}
}

@article{starikov2024angular,
  title={Angular-dependent interatomic potential for large-scale atomistic simulation of W-Mo-Nb ternary alloys},
  author={Starikov, Sergei and Grigorev, Petr and Olsson, P{\"a}r AT},
  journal={Computational Materials Science},
  volume={233},
  pages={112734},
  year={2024},
  publisher={Elsevier}
}

@article{starikov2025dislocation,
  title={Dislocation mobility function as a key to understanding plasticity of refractory metals and alloys},
  author={Starikov, Sergei},
  journal={Computational Materials Science},
  volume={246},
  pages={113411},
  year={2025},
  publisher={Elsevier}
}

@article{li2020complex,
  title={Complex strengthening mechanisms in the NbMoTaW multi-principal element alloy},
  author={Li, Xiang-Guo and Chen, Chi and Zheng, Hui and Zuo, Yunxing and Ong, Shyue Ping},
  journal={npj Computational Materials},
  volume={6},
  number={1},
  pages={70},
  year={2020},
  publisher={Nature Publishing Group UK London}
}

@article{ashby1985influence,
  title={The influence of dislocation density on the ductile-brittle transition in BCC metals},
  author={Ashby, MF and Embury, JD},
  journal={Scripta metallurgica},
  volume={19},
  number={4},
  pages={557--562},
  year={1985},
  publisher={Elsevier}
}

@article{johnson1962ductile,
  title={The ductile—brittle transition in body-centred cubic transition metals},
  author={Johnson, AA},
  journal={Philosophical Magazine},
  volume={7},
  number={74},
  pages={177--196},
  year={1962},
  publisher={Taylor \& Francis}
}

@article{han2022mechanism,
  title={Mechanism of ductile-to-brittle transition in body-centered-cubic metals: A brief review},
  author={HAN, Weizhong and LU, Yan and ZHANG, Yuheng},
  journal={Acta Metall Sin},
  volume={59},
  number={3},
  pages={335--348},
  year={2022}
}

@article{caillard2003thermally,
  title={Thermally activated mechanisms in crystal plasticity},
  author={Caillard, Daniel and Martin, JL and others},
  journal={MRS Bulletin},
  volume={30},
  pages={318--320},
  year={2003}
}

@book{argon2007strengthening,
  title={Strengthening mechanisms in crystal plasticity},
  author={Argon, Ali},
  volume={4},
  year={2007},
  publisher={OUP Oxford}
}

@article{duesbery1991dislocation,
  title={The dislocation core in crystalline materials},
  author={Duesbery, MS and Richardson, GY},
  journal={Critical Reviews in Solid State and Material Sciences},
  volume={17},
  number={1},
  pages={1--46},
  year={1991},
  publisher={Taylor \& Francis}
}

@article{christian1983some,
  title={Some surprising features of the plastic deformation of body-centered cubic metals and alloys},
  author={Christian, JW},
  journal={Metallurgical transactions A},
  volume={14},
  number={7},
  pages={1237--1256},
  year={1983},
  publisher={Springer}
}

@article{vitek1970core,
  title={The core structure of $1/2$ (111) screw dislocations in bcc crystals},
  author={Vitek, V and Perrin, RC and Bowen, DK},
  journal={Philosophical Magazine},
  volume={21},
  number={173},
  pages={1049--1073},
  year={1970},
  publisher={Taylor \& Francis}
}

@article{ito2001atomistic,
  title={Atomistic study of non-Schmid effects in the plastic yielding of bcc metals},
  author={Ito, K and Vitek, V},
  journal={Philosophical Magazine A},
  volume={81},
  number={5},
  pages={1387--1407},
  year={2001},
  publisher={Taylor \& Francis}
}

@article{wen2000atomistic,
  title={Atomistic simulation of kink-pairs of screw dislocations in body-centred cubic iron},
  author={Wen, M and Ngan, AHW},
  journal={Acta materialia},
  volume={48},
  number={17},
  pages={4255--4265},
  year={2000},
  publisher={Elsevier}
}

@article{gordon2010screw,
  title={Screw dislocation mobility in BCC metals: the role of the compact core on double-kink nucleation},
  author={Gordon, PA and Neeraj, T and Li, Y and Li, J},
  journal={Modelling and Simulation in Materials Science and Engineering},
  volume={18},
  number={8},
  pages={085008},
  year={2010}
}

@article{cereceda2013assessment,
  title={Assessment of interatomic potentials for atomistic analysis of static and dynamic properties of screw dislocations in W},
  author={Cereceda, David and Stukowski, Alexander and Gilbert, MR and Queyreau, Sylvain and Ventelon, Lisa and Marinica, Mihai-Cosmin and Perlado, JM and Marian, Jaime},
  journal={Journal of Physics: Condensed Matter},
  volume={25},
  number={8},
  pages={085702},
  year={2013},
  publisher={IOP Publishing}
}

@article{jiang2026scaling,
  title={Scaling reliable interatomic potentials to complex nuclear alloys via pretrained atomic models},
  author={Jiang, Mingxuan and Xu, Biao and Deng, Yixin and Ma, Shihua and Kai, Ji-Jung and Gao, Fei and Deng, Huiqiu},
  journal={npj Computational Materials},
  year={2026},
  publisher={Nature Publishing Group UK London}
}

@article{luo2026dislocation,
  title={Dislocation-induced ordering as a source of strengthening in refractory multi-principal element alloys},
  author={Luo, Yuhao and Wang, Tianyi and Huang, Zhihao and Su, Yanqing and Xu, Shuozhi and Liaw, Peter K and Li, Xiang-Guo},
  journal={npj Computational Materials},
  year={2026},
  publisher={Nature Publishing Group UK London}
}

@article{lysogorskiy2023active,
  title={Active learning strategies for atomic cluster expansion models},
  author={Lysogorskiy, Yury and Bochkarev, Anton and Mrovec, Matous and Drautz, Ralf},
  journal={Physical Review Materials},
  volume={7},
  number={4},
  pages={043801},
  year={2023},
  publisher={APS}
}

@article{vineyard1957frequency,
  title={Frequency factors and isotope effects in solid state rate processes},
  author={Vineyard, George H},
  journal={Journal of Physics and Chemistry of Solids},
  volume={3},
  number={1-2},
  pages={121--127},
  year={1957},
  publisher={Elsevier}
}

@article{dorn1963nucleation,
  title={Nuclearion of kink pairs and the Peierls' mechanism of plastic deformation},
  author={Dorn, John E and Rajnak, Stanley},
  journal={Transactions of the Metallurgical Society of AIME},
  volume={230},
  pages={1052--1064},
  year={1963}
}

@article{po2016phenomenological,
  title={A phenomenological dislocation mobility law for bcc metals},
  author={Po, Giacomo and Cui, Yinan and Rivera, David and Cereceda, David and Swinburne, Tom D and Marian, Jaime and Ghoniem, Nasr},
  journal={Acta Materialia},
  volume={119},
  pages={123--135},
  year={2016},
  publisher={Elsevier}
}

@article{zotov2022entropy,
  title={Entropy of kink pair formation on screw dislocations: an accelerated molecular dynamics study},
  author={Zotov, Nikolay and Grabowski, Blazej},
  journal={Modelling and Simulation in Materials Science and Engineering},
  volume={30},
  number={6},
  pages={065004},
  year={2022},
  publisher={IOP Publishing}
}

@article{yang2001kink,
  title={Kink-pair mechanisms for a/2 $\langle 111 \rangle$ screw dislocation motion in bcc tantalum},
  author={Yang, LH and Moriarty, JA},
  journal={Materials Science and Engineering: A},
  volume={319},
  pages={124--129},
  year={2001},
  publisher={Elsevier}
}

@article{kocks1975models,
  title={Models for macroscopic slip},
  author={Kocks, UF and Argon, AS and Ashby, MF},
  journal={Progress in Material Science},
  volume={19},
  pages={171--229},
  year={1975}
}

@article{marinica2013interatomic,
  title={Interatomic potentials for modelling radiation defects and dislocations in tungsten},
  author={Marinica, Mihai-Cosmin and Ventelon, Lisa and Gilbert, Mark R and Proville, Laurent and Dudarev, Sergei L and Marian, Jaime and Bencteux, Guy and Willaime, Fran{\c{c}}ois},
  journal={Journal of Physics: Condensed Matter},
  volume={25},
  number={39},
  pages={395502},
  year={2013},
  publisher={IOP Publishing}
}

@article{dezerald2015first,
  title={First-principles prediction of kink-pair activation enthalpy on screw dislocations in bcc transition metals: V, Nb, Ta, Mo, W, and Fe},
  author={Dezerald, L and Proville, L and Ventelon, Lisa and Willaime, F and Rodney, David},
  journal={Physical Review B},
  volume={91},
  number={9},
  pages={094105},
  year={2015},
  publisher={APS}
}

@article{proville2012quantum,
  title={Quantum effect on thermally activated glide of dislocations},
  author={Proville, Laurent and Rodney, David and Marinica, Mihai-Cosmin},
  journal={Nature materials},
  volume={11},
  number={10},
  pages={845--849},
  year={2012},
  publisher={Nature Publishing Group UK London}
}

@article{vitek2004influence,
  title={Influence of non-glide stresses on plastic flow: from atomistic to continuum modeling},
  author={Vitek, V and Mrovec, M and Bassani, JL},
  journal={Materials Science and Engineering: A},
  volume={365},
  number={1-2},
  pages={31--37},
  year={2004},
  publisher={Elsevier}
}

@article{groger2008multiscale,
  title={Multiscale modeling of plastic deformation of molybdenum and tungsten: I. Atomistic studies of the core structure and glide of $1/2\langle 111\rangle$ screw dislocations at 0 K},
  author={Gr{\"o}ger, R and Bailey, AG and Vitek, V},
  journal={Acta Materialia},
  volume={56},
  number={19},
  pages={5401--5411},
  year={2008},
  publisher={Elsevier}
}

@article{Gordon_2011,
author = { P.A.   Gordon  and  T.   Neeraj  and  M.I.   Mendelev },
title = {Screw dislocation mobility in BCC Metals: a refined potential description for {$\alpha$}-Fe},
journal = {Philosophical Magazine},
volume = {91},
number = {30},
pages = {3931-3945},
year  = {2011},
publisher = {Taylor & Francis},
doi = {10.1080/14786435.2011.597947},
}

@article{rodney2009stress,
  title={Stress-dependent Peierls potential: Influence on kink-pair activation},
  author={Rodney, David and Proville, Laurent},
  journal={Physical Review B},
  volume={79},
  number={9},
  pages={094108},
  year={2009},
  publisher={APS}
}

@article{starikov2021optimized,
  title={Optimized interatomic potential for atomistic simulation of Zr-Nb alloy},
  author={Starikov, S and Smirnova, D},
  journal={Computational Materials Science},
  volume={197},
  pages={110581},
  year={2021},
  publisher={Elsevier}
}

@article{leveau2025segregation,
  title={Segregation of hydrogen on screw dislocations in tungsten and its impact on dislocation mobility},
  author={Leveau, Thomas and Ventelon, Lisa and Marinica, Mihai-Cosmin and Clouet, Emmanuel},
  journal={Acta Materialia},
  pages={121869},
  year={2025},
  publisher={Elsevier}
}

@article{allera2025activation,
  title={Activation entropy of dislocation glide in body-centered cubic metals from atomistic simulations},
  author={Allera, Arnaud and Swinburne, Thomas D and Goryaeva, Alexandra M and Bienvenu, Baptiste and Ribeiro, Fabienne and Perez, Michel and Marinica, Mihai-Cosmin and Rodney, David},
  journal={Nature Communications},
  volume={16},
  number={1},
  pages={8367},
  year={2025},
  publisher={Nature Publishing Group UK London}
}

@article{chaari2014first,
  title={First-principles study of secondary slip in zirconium},
  author={Chaari, Nermine and Clouet, Emmanuel and Rodney, David},
  journal={Physical review letters},
  volume={112},
  number={7},
  pages={075504},
  year={2014},
  publisher={APS}
}

@article{zotov2021molecular,
  title={Molecular dynamics simulations of screw dislocation mobility in bcc Nb},
  author={Zotov, Nikolay and Grabowski, Blazej},
  journal={Modelling and Simulation in Materials Science and Engineering},
  volume={29},
  number={8},
  pages={085007},
  year={2021},
  publisher={IOP Publishing}
}

@article{podryabinkin2017active,
  title={Active learning of linearly parametrized interatomic potentials},
  author={Podryabinkin, Evgeny V and Shapeev, Alexander V},
  journal={Computational Materials Science},
  volume={140},
  pages={171--180},
  year={2017},
  publisher={Elsevier}
}

@article{wang2024taming,
  title={The taming of the screw: Dislocation cores in BCC metals and alloys},
  author={Wang, Rui and Zhu, Lingyu and Pattamatta, Subrahmanyam and Srolovitz, David J and Wu, Zhaoxuan},
  journal={Materials Today},
  volume={79},
  pages={36--48},
  year={2024},
  publisher={Elsevier}
}

@article{byggmastar2021modeling,
  title={Modeling refractory high-entropy alloys with efficient machine-learned interatomic potentials: Defects and segregation},
  author={Byggm{\"a}star, Jesper and Nordlund, Kai and Djurabekova, Flyura},
  journal={Physical Review B},
  volume={104},
  number={10},
  pages={104101},
  year={2021},
  publisher={APS}
}

@article{aitken2024controlling,
  title={Controlling screw dislocation core structure and Peierls barrier in BCC interatomic potentials},
  author={Aitken, Zachary H and Sorkin, Viacheslav and Yu, Zhi Gen and Chen, Shuai and Tan, Teck Leong and Wu, Zhaoxuan and Zhang, Yong-Wei},
  journal={International Journal of Solids and Structures},
  volume={303},
  pages={113004},
  year={2024},
  publisher={Elsevier}
}

@article{sobol1967distribution,   
    title={Distribution of points in a cube and approximate evaluation of integrals},   
    author={Sobol, Ilya M},   
    journal={USSR Computational Mathematics and Mathematical Physics},   
    volume={7},   
    pages={86--112},   
    year={1967} 
}

\end{document}